\documentclass[twocolumn]{aastex631}

\usepackage{amsmath}
\usepackage{float}
\usepackage[section]{placeins}
\usepackage{multirow}
\graphicspath{{./}{}}
\shorttitle{The faint end of the galaxy stellar mass function at $z\simeq 4-8$}
\shortauthors{Navarro-Carrera et al.}

\begin{document}

\title{Constraints on the Faint End of the Galaxy Stellar Mass Function at $z\simeq4-8$ from Deep \textit{JWST} Data}

\newcommand{\gsim}{\raisebox{-0.13cm}{~\shortstack{$>$ \\[-0.07cm] $\sim$}}~}
\newcommand{\lsim}{\raisebox{-0.13cm}{~\shortstack{$<$ \\[-0.07cm] $\sim$}}~}

\correspondingauthor{Rafael Navarro-Carrera}

\affiliation{Kapteyn Astronomical Institute, University of Groningen, P.O. Box 800, 9700AV Groningen, The Netherlands}
\email{navarro@astro.rug.nl}
\author[0000-0001-6066-4624]{Rafael Navarro-Carrera}

\affiliation{Kapteyn Astronomical Institute, University of Groningen, P.O. Box 800, 9700AV Groningen, The Netherlands}

\author[0000-0002-5104-8245]{Pierluigi Rinaldi}
\affiliation{Kapteyn Astronomical Institute, University of Groningen, P.O. Box 800, 9700AV Groningen, The Netherlands}

\author[0000-0001-8183-1460]{Karina I. Caputi}
\affiliation{Kapteyn Astronomical Institute, University of Groningen, P.O. Box 800, 9700AV Groningen, The Netherlands}
\affiliation{Cosmic Dawn Center (DAWN), Copenhagen, Denmark
}

\author[0000-0001-8386-3546]{Edoardo Iani}
\affiliation{Kapteyn Astronomical Institute, University of Groningen, P.O. Box 800, 9700AV Groningen, The Netherlands}

\author[0000-0002-5588-9156]{Vasily Kokorev}
\affiliation{Kapteyn Astronomical Institute, University of Groningen, P.O. Box 800, 9700AV Groningen, The Netherlands}

\author[0000-0001-8289-2863]{Sophie van Mierlo}
\affiliation{Kapteyn Astronomical Institute, University of Groningen, P.O. Box 800, 9700AV Groningen, The Netherlands}



\begin{abstract}
We analyze a sample of 3300 galaxies between redshifts $z\simeq 3.5$ and $z\simeq 8.5$ selected from \textit{JWST} images in the Hubble Ultra Deep Field (HUDF) and UKIDSS Ultra Deep Survey field, including objects with stellar masses as low as $\simeq 10^8 \, \rm{{\rm \ M_\odot}}$ up to $z\simeq 8$. The depth and wavelength coverage of the \textit{JWST} data allow us, for the first time, to derive robust stellar masses for such high-$z$, low stellar-mass galaxies on an individual basis. We compute the galaxy stellar mass function (GSMF), after complementing our sample with ancillary data from CANDELS to constrain the GMSF at high stellar masses ($\mathcal{M}>M^\ast$). Our results show a steepening of the low stellar-mass end slope ($\alpha$) with redshift, with $\alpha = -1.61\pm0.05$ at $z\simeq 4$ and $\alpha=-1.98\pm0.14$ at $z\simeq 7$. We also observe an evolution of the normalization $\phi^\ast$ from $z\simeq 7$ to $z\simeq 4$, with $\phi^\ast_{z\simeq 4}/\phi^\ast_{z\simeq 7} = 130^{+210}_{-50}$. Our study incorporates a novel method for the estimation of the Eddington bias that takes into account its possible dependence both on stellar mass and redshift, while allowing for skewness in the error distribution. We finally compute the resulting cosmic stellar mass density and find a flatter evolution with redshift than previous studies. 
\end{abstract}



\section{INTRODUCTION} \label{sec:intro}

Over the past decades, substantial progress has been made in our understanding of when and how galaxies have assembled their stellar masses. The redshift and stellar-mass ranges that could be studied have become steadily larger,  thanks to the progressive availability of powerful galaxy surveys that probe increasingly wider areas and/or provide deeper photometric coverage (e.g., \citealp{ilbert_mass_2013,  caputi_spitzer_2015, grazian_galaxy_2015, davidzon_cosmos2015_2017, deshmukh_spitzer_2018}). 

However, until now, the study of low stellar-mass galaxies ($M \lsim 10^8 - 10^9 \, \rm {\rm \ M_\odot}$) at high redshifts has been very difficult, as the existing galaxy surveys were not deep enough to enable the investigation of this region in parameter space. A notable exception were massive galaxy cluster fields, where the lensing magnification helped to detect intrinsically faint, distant objects  in the background, which  could not be seen in blank-field surveys of comparable depths \citep[e.g.,][]{karman_muse_2017,bhatawdekar_evolution_2019,kikuchihara_early_2020,vanzella_muse_2021, santini_stellar_2022}.  Nonetheless, sufficiently deep imaging to detect these objects was, in most cases, only available with the \textit{Hubble Space Telescope (HST)} up to $\simeq 1.6 \, \rm \mu m$. This implied that, for galaxies at $z \gsim 3$, only the rest-UV light could be traced and any derived stellar-mass estimate relied on long extrapolations of the galaxy spectral models.

The advent of \textit{JWST} is now radically changing this situation. By providing high-sensitivity imaging up to mid-infrared wavelengths, \textit{JWST} is allowing us for the first time to find the building blocks for galaxy formation at high redshifts and directly measure the light emitted by their evolved stellar populations, which is necessary to obtain robust estimates of their stellar masses. Therefore, we can now investigate early stellar mass assembly down to the typical stellar masses of local satellite galaxies. This is the main goal of this work.

The galaxy stellar mass function (GSMF) is one of the most important statistical tools to understand galaxy evolution across cosmic time.  The evolution of this function with redshift reveals how galaxies of different stellar masses have assembled until producing the final distribution that we see in the Universe today.  A Schechter function \citep{schechter_analytic_1976} is commonly used as a parametrization for the GSMF. This function is characterized by a power-law shape with slope $\alpha$ (low-mass end slope) up to a characteristic stellar mass ($\mathcal{M}$), and an exponential decline at the high-mass end.  The GSMF constitutes an important statistical tool to understand galaxy formation  and reproducing the GSMF at different redshifts is one of the main challenges of galaxy formation models. Indeed a number of studies have provided strong constraints on the GSMF at low redshifts (e.g., \citealp{pozzetti_vimos_2007, ilbert_galaxy_2010, wright_galaxy_2017, bates_mass_2019}).  Some of these works have also investigated the GSMF's dependence on galaxy types, such as star-forming/quiescent galaxies (e.g., \citealp{davidzon_cosmos2015_2017}), central/satellite galaxies (e.g.,\citealp{yang_galaxy_2009}), and galaxy morphology \citep{ilbert_galaxy_2010}.  Moreover, the evolution of the GSMF with cosmic time provides information on the different physical processes that shape galaxy assembly  (e.g., supernova and super massive black-hole feedback and mergers) and clues to understand star-formation and quenching mechanisms in galaxies of different stellar masses (e.g., \citealp{peng_mass_2010, weaver_cosmos2020_2022}).

In particular, the value of the GSMF faint-end slope is a crucial constraint for galaxy formation models. Generally, these models predict a steepening of this low-mass end with redshift, meaning that low stellar-mass galaxies were relatively more abundant at earlier cosmic times. However, for the reasons explained above, determining observationally the GSMF low-mass end at high redshifts has been very challenging. A number of studies have attempted to constrain the GSMF faint-end slope, both from deep blank and lensing fields, but no consensus has been reached yet on its value and redshift evolution \citep[e.g.,][]{grazian_galaxy_2015, song_evolution_2016, bhatawdekar_evolution_2019, kikuchihara_early_2020}. With ultra-deep \textit{JWST} imaging in hand, we can now study the GSMF faint-end slope at high redshift in a much more robust way than ever before.

In its first year of operations, \textit{JWST} has been able to study unknown or poorly constrained galaxy populations due to its unprecedented near and mid infrared sensitivity. Several studies are making use of these \textit{JWST} unique capabilities to probe different aspects of galaxy evolution at high redshifts,  both from the analysis of normal galaxies (e.g., \citealp{bagley_ceers_2022, iani_first_2022, finkelstein_ceers_2023, jin_massive_2023, kokorev_jwst_2023, rinaldi_strong_2023, perez-gonzalez_life_2023}), as well as active galactic nuclei \citep[e.g.,][]{yang_ceers_2023}. 

In this work, we make use of  very deep \textit{JWST} Near Infrared Camera (NIRCam) observations to study the evolution of the GSMF (focusing on the low-mass end) at high redshifts up to the Epoch of Reionization, between $z\simeq 3.5$ and $z\simeq 8.5$.  We consider the publicly available NIRCam images of the UKIDSS Ultra Deep Survey (UDS; \citealp{lawrence_ukirt_2007}) and Hubble Ultra Deep Field (HUDF;  \citealp{beckwith_hubble_2006}; \citealp{illingworth_hst_2013}) fields for our analysis. We study a sample of $\simeq 3300$ galaxies between $z\simeq 3.5$ and $z\simeq 8.5$, using for first time \textit{JWST} data to compute the GSMF, focusing our analysis on the evolution of the low-mass end slope with cosmic time and the derived  evolution of the cosmic stellar mass density. 

This paper is structured as follows: in Sect. \ref{sec:dataset} we describe the dataset and in Sect. \ref{sec:dataset_sedfitting} the SED fitting procedure and estimation of photometric redshifts and stellar masses. The methodology used for computing the GSMF is presented in Sect. \ref{sec:method}, followed by the results on the GSMF evolution in Sect. \ref{sec:smf}.  Finally, we investigate the evolution of the low-mass end slope and cosmic stellar mass density in Sect. \ref{sec:smf_alpha_evol} and Sect. \ref{sec:stellar_mass_density}, respectively. We adopt a cosmology with $\Omega_{m,0}=0.3$, $\Omega_{\Lambda,0}=0.7$, and $H_0 = $ 70 km s$^{-1}$ Mpc$^{-1}$. Magnitudes are given in the AB system \citep{oke_secondary_1983}. Densities of the galaxies are measured on comoving scales. We adopt the \cite{chabrier_galactic_2003} initial mass function (IMF) in a stellar mass range of $\mathcal{M}:\rm{0.1\simeq 100} \, {\rm \ M_\odot}$ to estimate stellar masses. In this paper, all of the stellar masses taken from the previous studies are converted to those estimated with the \cite{chabrier_galactic_2003} IMF.

\section{DATASETS} \label{sec:dataset}
\subsection{General overview}
\begin{figure*}[ht]
\centering
\includegraphics[width=1.85\columnwidth]{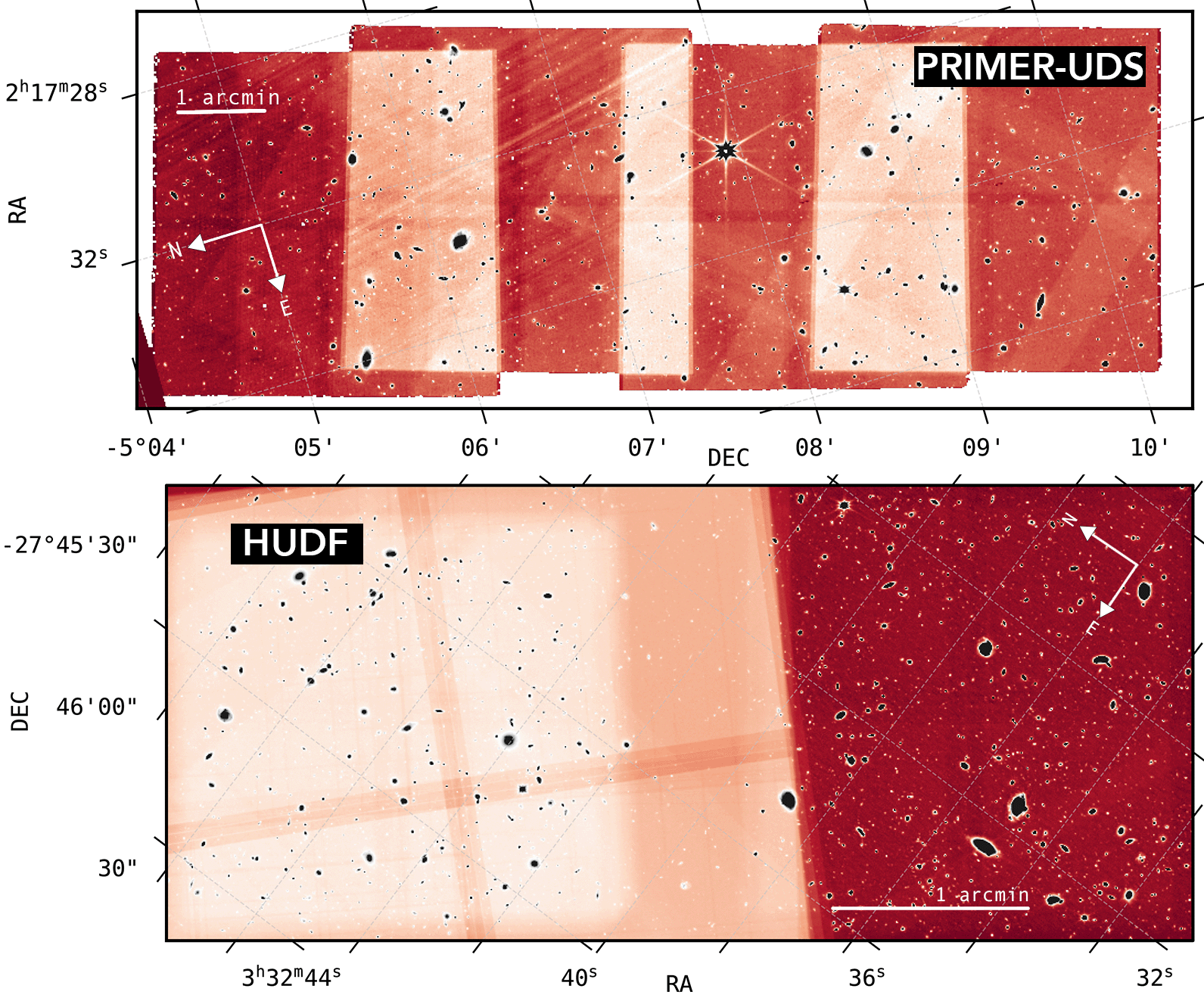}
\caption{\textbf{Top:} PRIMER-UDS field. \textbf{Bottom:} \textit{JWST} HUDF field. Sources are marked in black and background in red. Lighter shades indicate deeper areas. The XDF can be seen as the lightest patch on the left-hand side of the field.}
\label{fig:fields}
\vspace{0 mm}
\end{figure*}

In this work we  make use of photometric datasets from two different fields partly covered with \textit{JWST}:  HUDF ($\simeq12$ arcmin$^2$) and UDS ($\simeq7$ arcmin$^2$). The study of two different fields helps to mitigate the effects of cosmic variance, while providing a statistically larger galaxy sample populating the low and intermediate stellar-mass regimes. The region of the HUDF with \textit{JWST} coverage includes the entire Hubble eXtreme Deep Field (XDF, \citealp{illingworth_hst_2013}), which benefits from the deepest \textit{HST} observations (typical depth of $30$ ABMag at $5\sigma$) and some of the deepest public \textit{JWST} imaging. For both fields, we also make use of ancillary \textit{HST} data to obtain our galaxy catalogs, as explained below.

Albeit very useful for the low stellar-mass end, the \textit{JWST} fields considered here are too small to properly constrain the GSMF around the turnover stellar mass $M^\ast$. Therefore, we complement our galaxy catalogs in the \textit{JWST}-covered fields with the publicly available source catalogs from  the Cosmic Assembly Near-Infrared Deep Extragalactic Legacy Survey  \citep[CANDELS;][]{grogin_candels_2011, koekemoer_candels_2011} in the GOODS-S field ($\simeq170$ arcmin$^2$) \citep{galametz_candels_2013} and the UDS field ($\simeq200$ arcmin$^2$) \citep{guo_candels_2013}. The larger area covered by CANDELS allows us to design a tiered strategy,  with very deep \textit{JWST} observations to probe the faint end (in small areas) and wider-area data to constrain the GSMF at the high-mass end.

\subsubsection{HUDF}
\textit{JWST} data for HUDF (including XDF) have been obtained by \cite{williams_jems_2023}, as part of a General Observers Cycle-1 program (Proposal-ID: 1963; PI: Christina C. Williams). The obtained datasets consist of five \textit{JWST}-NIRCam medium bands: F182M, F210M, F430M, F460M, and F480M. We also make use of NIRCam data obtained by \textit{The First Reionization Epoch Spectroscopic COmplete Survey} (FRESCO; \citealp{oesch_jwst_2023}) (Proposal-ID: 1895; PI: Pascal Oesch), which provides complementary F444W imaging and additional exposure time in bands F182M ad F210M, in the same field.

In our study we include both the XDF and nearby region of the HUDF field that lies within the \textit{JWST} pointings. Two regions with different average depth have been defined: deep (XDF) and shallow (the rest of the field). The depths and areas are summarized in Fig. \ref{fig:fields} and presented in Table \ref{tab:fields}.

Our procedure for the NIRCam data reduction is the same as the one described in \cite{rinaldi_strong_2023}. The version of the \textit{JWST} pipeline used for is 1.8.2 and the Calibration Reference Data System (CRDS) pipeline mapping (pmap) version is 1018. We have added extra steps to the original pipeline provided by the Space Telescope Science Institute \citep[STScI; ][]{bushouse_jwst_2020},  following a similar approach to the one of \cite{bagley_ceers_2022}. These new steps include correction for stripping (\textit{1/f} noise), stray-light effects (wisps) and residual cosmic rays (snow balls). As mentioned in \cite{rinaldi_strong_2023}, photometry has been checked for consistency against the default pipeline without any additional steps.

All the images have been resampled to pixel scale of $\rm 0\farcs{03}/pix$ and drizzled to a mosaic using the Hubble Legacy Fields catalog as reference to align the images. 

As a complement for the \textit{JWST} images, we also make use of \textit{HST} ancillary imaging over the HUDF, which has been obtained from the Hubble Legacy Field GOODS-S program (HLF-GOODS-S; \citealp{whitaker_hubble_2019}), covering a wide spectral range: $0.2 \mu m - 1.6 \mu m$. A total of 13 photometric bands are available, as follows.  In the ultra-violet WFC3/UVIS F225W, F275W, F336W; in  the optical: ACS/WFC F435W, F606W, F775W, F814W, F850LP; and in the near-infrared: WFC3/IR F098M, F105W, F125W, F140W and F160W.

\subsubsection{UDS}
\textit{JWST} data in part of the UDS field has been obtained by the \textit{Public Release IMaging for Extragalactic Research} (PRIMER; \citealp{dunlop_primer_2021}), a General Observers Cycle-1 program  (Proposal-ID: 1837; PI: James Dunlop). The PRIMER dataset consists of imaging in 8 NIRCam bands: F090W, F115W, F150W, F200W, F277W, F356W, F410M and F444W.

Our procedure for the  NIRCam data reduction is similar to that for the HUDF. For the UDS, however, due to presence of stray light in some of the filters we need to apply a mask to exclude regions of severely contaminated photometry. The reduction in area after the masking is significant, but less than 50\%, accounting for the loss of all the detector area affected by the stripes. The effective area of the UDS-PRIMER dataset is then $\simeq 7$ arcmin$^2$.

As a result of overlapping exposures within the field, the depth is non-uniform. We have identified two regions: deep and shallow. The area and depth in each band \textit{JWST} can be found in Table \ref{tab:fields} and the map of the field is presented in Fig. \ref{fig:fields}.

The \textit{HST} ancillary imaging is obtained from CANDELS-UDS data \citep{galametz_candels_2013}, including newer observations with \textit{HST/WFC3} that where not present in the original \cite{galametz_candels_2013} analysis. The coverage is $0.4 \mu m - 1.6 \mu m$, with 7 photometric bands. In the optical: ACS/WFC F435W, F606W, F814W. In the near-infrared: WFC3/IR, F105W, F125W, F140W and F160W.

\begin{deluxetable}{l|cccc}[]
\tabcolsep=2mm
\tablecaption{\label{tab:fields} Area and depth (5$\sigma$ limiting magnitudes measured in apertures of $\rm 0\farcs{25}/pix$ radius) for \textit{JWST} fields used for our analysis. The depths in \textit{JWST} bands is uniform in the case of the HUDF, while the \textit{HST} data has different depths in the XDF and the rest of HUDF regions.}
\startdata
 & & & &\\
 \textbf{area} &  \textbf{UDS}  &  \textbf{UDS} & \textbf{HUDF} \\
  (arcmin$^2$) &  shallow & deep &  7.16 \textit{HST}-deep\\
 & 2.35 & 4.60 & 5.00 \textit{HST}-shallow\\
\hline
$5\sigma$ \textbf{lim. mag}    &    &       & \\
F090W & 27.25 & 27.60 & -\\
F115W & 27.35 & 27.70 & -\\
F150W & 27.55 & 27.75 & -\\
F182M & - & - & 29.45\\
F200W & 27.70 & 27.90 & -\\
F210M & - & - & 29.30\\
F277W & 28.18 & 28.25 & -\\
F356W & 28.20 & 28.22 & -\\
F410M & 27.48 & 27.65& -\\
F430M & - & - & 28.40\\
F444W & 27.80 & 27.95 & 28.10\\
F460M & - & - & 28.15\\
F480M & - & - & 28.30\\
\enddata
\end{deluxetable}

\section{PHOTOMETRIC CATALOG AND SED FITTING} \label{sec:dataset_sedfitting}
\subsection{\rm{Photometric catalog}}\label{sec:catalogs}
We have detected sources and measured photometry in both the HUDF and UDS field regions with \textit{JWST} coverage,   making use of the software Source Extractor (\textsc{SExtractor}, \citealp{bertin_sextractor_1996}). We considered a stacked image that combines all the available \textit{JWST} bands as detection map, and then ran \textsc{SExtractor} in dual-image mode to measure photometry in every band.

As parameter values for source dection, we used those corresponding to the \texttt{hot-mode} described in \cite{galametz_candels_2013}, optimized for the detection of faint sources, which is ideal for low stellar-mass galaxies at high redshift.

We measured source photometry following \cite{rinaldi_strong_2023}, i.e., in fixed $\rm 0\farcs{5}$ diameter circular apertures and Kron apertures \citep[i.e., \texttt{flux\_auto};][]{kron_photometry_1980}. We corrected the aperture photometry to total considering the curve of growth of the corresponding point spread function in each filter, using \textsc{WebbPSF} \citep{perrin_updated_2014}.  For each source with apparent magnitude $<27$ (as described in \cite{rinaldi_strong_2023}), we chose the brightest flux (either the corrected aperture flux or \texttt{flux\_auto}). For all fainter sources, which can be considered point-like even in \textit{JWST} images, we adopted the corrected aperture fluxes for our photometric catalog.

We corrected all source fluxes for Galactic extinction with the help of \texttt{dustmaps} \citep{green_dustmaps_2018}, which is in good agreement with the prescriptions from \cite{schlafly_measuring_2011} for \textit{HST} filters. To take into account possible flux calibration uncertainties and because \textsc{SExtractor} has been found to underestimate photometric errors (e.g., \citealp{sonnett_testing_2013}), we have imposed a minimum error of $0.05$~mag to all the photometry.

Flux upper limits have been estimated for all sources with image coverage, but not detected, in any given band, by measuring the local background r.m.s from multiple, circular apertures of $\rm 0\farcs{5}$ diameter, randomly positioned around each source. In the case that no photometric coverage is available in a given band (because the source is not contained in the field of view), we ignored that band for the SED fitting explained below.

Our source catalogs have been cleaned for Galactic stars by cross-matching our catalog with \textit{Gaia} Data Release 3 \citep{gaia_collaboration_gaia_2022} and by excluding all sources with high stellarity parameter (i.e., \texttt{CLASS\_STAR} $>$ 0.8) that lie on the stellar locus of the (F435W $-$ F125W) vs (F125W $-$ F444W) diagram, following the same criterion as in \cite{caputi_stellar_2011}.

\begin{deluxetable*}{ccchlDlc}[ht!]
\tablecaption{SED parameters employed for the \textsc{LePHARE} run, both for \texttt{BC03} and \texttt{SB99} stellar population models used in this work.}
\label{tab:sedfit_params}
\tablewidth{0pt}
\tablehead{
\colhead{Parameter} & \colhead{\textbf{BC03}} & \colhead{\textbf{SB99}$^1$}
}
\startdata
Templates & \citet{bruzual_stellar_2003} & \citet{leitherer_starburst99_1999} \\ $e-$folding time ($\tau$) & 0.01 - 15 (8 steps) + SSP & constant SFH\\
Metallicty (Z/Z$_{\odot}$) & 0.02; 1.00  & 0.05; 0.40\\
Age (Gyr) & 0.001 - 13.5 (49 steps) & 0.001 - 0.1 (6 steps)\\
& \multicolumn{2}{c}{Common values$^2$}\\
Extinction laws & \multicolumn{2}{c}{\citet{calzetti_dust_2000} + \citet{leitherer_global_2002}}\\
E(B-V) & \multicolumn{2}{c}{ 0-1.5 (16 steps)}\\
IMF & \multicolumn{2}{c}{\citet{chabrier_galactic_2003}}\\
Redshift & \multicolumn{2}{c}{0-20 (291 steps)}\\
Emission lines & \multicolumn{2}{c}{Yes}\\
Cosmology (H$\mathrm{_{0}}$, $\mathrm{\Omega_{0}}$, $\mathrm{\Lambda_{0}}$) & \multicolumn{2}{c}{70, 0.3, 0.7}
\enddata
\tablecomments{$^1$In the case of \texttt{SB99} models, an identical configuration with respect to \texttt{BC03} for the extinction law, E(B-V), IMF, redshift interval, and cosmology has been used. However we opted for only 6 ages for the \texttt{SB99} run because nebular emission is only significant for very low age stellar populations. $^2$ The age grid is designed to provide a finer sampling for young ages.}
\end{deluxetable*}
\subsection{\rm{Photometric redshifts and stellar masses}} \label{sec:photometry}
\begin{figure}[h]
\includegraphics[width=1\columnwidth]{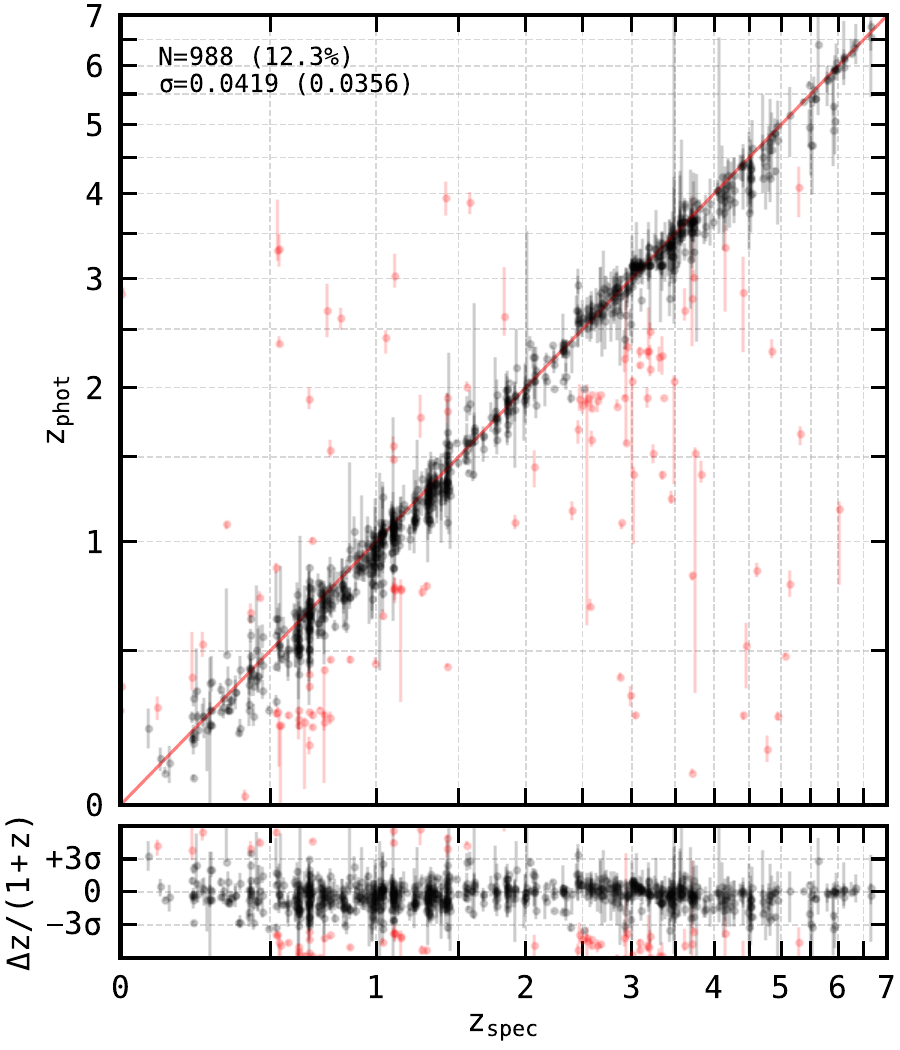}
\caption{Stellar mass and photometric redshift distribution of the galaxy sample used for our study. Catastrophic outliers are shown in red. The total number of galaxies in the sample is 988, out of which 12.3\% are catastrophic outliers. The median absolute deviation is $0.0419$, or $0.0356$ after removing the catastrophic outliers from the sample.}
\label{fig:zphot_zspec}
\end{figure}
\begin{figure}[t]
\includegraphics[width=1\columnwidth]{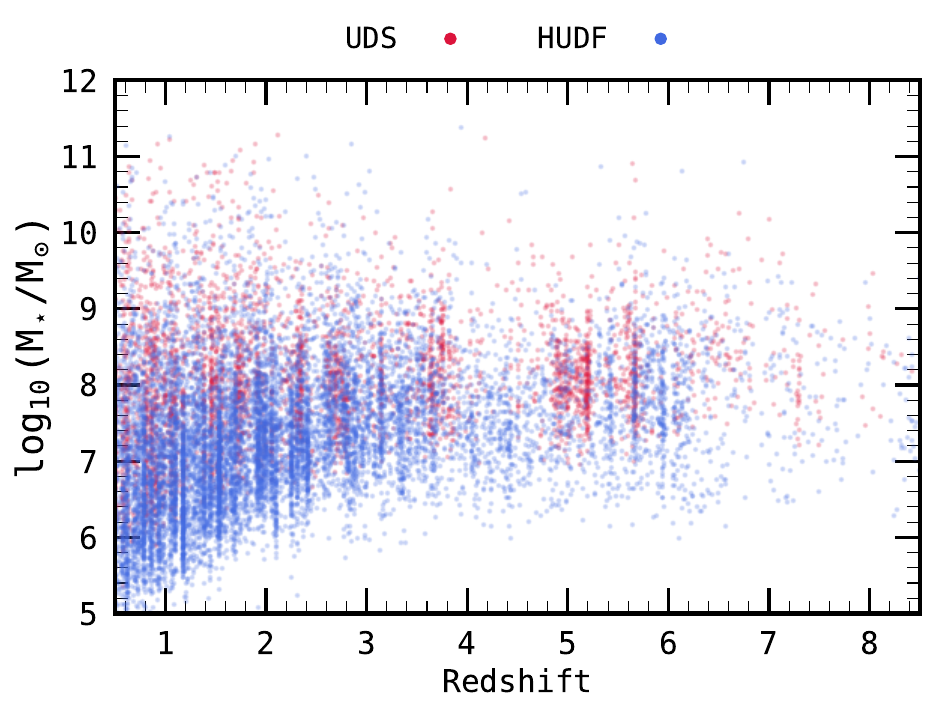}
\caption{Stellar mass and photometric redshift distribution of the galaxy sample used for our study. UDS and HUDF are shown as red and blue dots, respectively. From the plot, it can be seen that the difference between the limiting mass of HUDF and UDS is $1$ dex or larger, consistent with the deeper data that we use in HUDF.}
\label{fig:stellarmass_redshift}
\end{figure}
We have estimated the photometric redshifts and stellar masses of our galaxies  by performing SED fitting using the software \textsc{LePHARE} \citep{arnouts_lephare_2011}.  We considered a set of stellar population synthesis models from \citet[][hereafter \texttt{BC03}]{bruzual_stellar_2003}, including a single stellar population and a series of exponentially declining star formation histories with $\tau =$ 0.01, 0.1, 0.3, 1.0, 3.0, 5.0, 10.0, and 15 Gyr. All models have been constructed for two possible metallicities, namely solar ($\rm Z_\odot = 0.02$) and sub-solar ($\rm Z = 0.2 \times Z_\odot = 0.004$).

We also considered stellar population synthesis models from \citet[][hereafter \texttt{SB99}]{leitherer_starburst99_1999}  in our analysis. These additional models consist of a set of five young galaxy templates with ages spanning $\rm 10^6–10^8$ yr and constant star formation rates between $0.01$ $\rm {\rm \ M_\odot}$yr$^{-1}$ and $10$ ${\rm M_\odot}$yr$^{-1}$. All of them correspond to  sub-solar metallicities of $\rm Z = 0.008$ and $\rm Z = 0.001$. All the relevant paramters used for the SED fitting both for \texttt{BC03} and \texttt{SB99} can be found in Table \ref{tab:sedfit_params}.

These templates are characterized by incorporating the contribution of nebular continuum emission, especially relevant for young star-forming galaxies. Some previous work explores the effect produced by nebular emission in SED fitting using broadband photometry (e.g., \citealp{stark_keck_2013}, \citealp{de_barros_properties_2014}). The effect of nebular emission is to enhance the measured flux, resulting in a overestimation of stellar masses  \citep{bisigello_statistical_2019}, especially for young and low-mass galaxies. 

There have been previous attempts to study this effect (e.g., \citealp{stark_keck_2013, duncan_mass_2014, salmon_relation_2015, grazian_galaxy_2015}. \cite{stark_keck_2013} finds that the stellar masses obtained using SED fitting of \textit{Spitzer Space Telescope} (\citealp{werner_spitzer_2004}) data where overestimated by a factor of 2 to 4 when accounting for nebular line and continuum emission in redshifts $\rm z \simeq 6 - 7$. \cite{duncan_mass_2014} finds that stellar masses estimated including nebular emission are significantly lower. The models used for our work (\texttt{SB99}) only account for nebular continuum, so the effect on stellar masses is expected to be milder than the one including both contributions.

For all \texttt{BC03} models we considered a Chabrier \citep{chabrier_galactic_2003} initial mass function (IMF), and the \texttt{SB99} models we used were re-scaled to a Chabrier IMF. We convolved the spectral models with a reddening law following the prescription of \cite{calzetti_dust_2000} and \cite{leitherer_global_2002}, with color excess values of $0 \leq E(B-V) \leq 1.5$ equally distributed in steps of $0.1$ for both \texttt{BC03} and \texttt{SB99}.

We used our photometric catalog as the input catalog for SED fitting with \textsc{LePHARE}, considering  $3\sigma$ flux upper limits in all cases, but source non detection in the particular band. All templates that produce fluxes greater than the $3\sigma$ upper limits are rejected by \textsc{LePhare}. The adopted stellar mass is obtained from the set of models (\texttt{BC03} or \texttt{SB99}) that has the lowest value of $\chi^2$. The diagnostic of photometric versus spectroscopic redshift for the whole sample (Fig. \ref{fig:zphot_zspec}) shows that the fraction of catastrophic outliers (defined in our case as $(z_{phot}-z_{spec})/(1+z_{spec})>0.15$) is $12.3\%$, and the normalized median absolute deviation of the sample is $0.0419$.

Within our sample there are a total of 1804 galaxies that lie above the completeness limit (described in Sect. \ref{subsec:method_photometric_completeness}) in the redshift range $z\simeq 3.5$ to $z\simeq 8.5$. Table \ref{tab:smf_results} shows the number of galaxies used for computing the GSMF in each redshift interval. The resulting stellar mass versus redshift plot is shown in Fig. \ref{fig:stellarmass_redshift}.

\section{GSMF computation} \label{sec:method}

\subsection{\rm 1/Vmax method} \label{subsec:method_vmax}

We apply the 1/Vmax method \citep[][,described in more detail in Sect. \ref{subsec:method_vmax}]{schmidt_space_1968} to compute the GSMF. The 1/Vmax method is non-parametric and does not assume any particular functional form for the GSMF, but requires binning in stellar mass.  As shown by \cite{davidzon_cosmos2015_2017}, 1/Vmax provides good constraints to the GSMF, although other methods such as STY \citep{sandage_velocity_1979} are also commonly used in the literature (e.g., \citealp{caputi_spitzer_2015}).

To compute the GSMF with the 1/Vmax method, we binned our galaxy samples into redshift and stellar-mass bins.  The  corresponding comoving number density of galaxies for each stellar mass bin ($\mathcal{M}_0$, $\mathcal{M}_1$) can be determined using
\begin{equation}
\Phi(\mathcal{M})= \sum_{bin}\frac{1}{V_{max}} \times \frac{1}{\log_{10}(\mathcal{M}_1/{\rm M_\odot})-\log_{10}(\mathcal{M}_{0}/{\rm M_\odot})}\,,
\end{equation}
where $V_{max}$ is the comoving volume of each galaxy corrected in the following way:
\begin{equation}
V_{max} =   \left\{
\begin{array}{@{}l@{}}
	V^{bin}_{max} - V^{bin}_{min}  \ \  {\rm if} \  z_{lim} \geq z^{bin}_{max}\\
	V_{z_{lim}} - V^{bin}_{min} \ \ {\rm if} \  z_{lim} < z^{bin}_{max}
  \end{array}\right.\,.
\end{equation}
Here $V^{bin}_{min}$ and $V^{bin}_{max}$ are the comoving volumes corresponding to the redshift bin ($z^{bin}_{min}$, $z^{bin}_{max}$). $V_{z_{lim}}$ is the comoving volume at $z_{lim}$, the maximum (limiting) redshift at which each galaxy would lie within the flux completeness of the sample. The value of $z_{lim}$ for each galaxy can be obtained from the following relation between flux and redshift:
\begin{equation}
\frac{d^2(z_{max})}{1+z_{max}} = \frac{f_\nu^{obs}}{f_\nu^{lim}}\frac{d^2(z_{obs})}{1+z_{obs}}\,.
\end{equation}

\begin{figure*}[t]
\plottwo{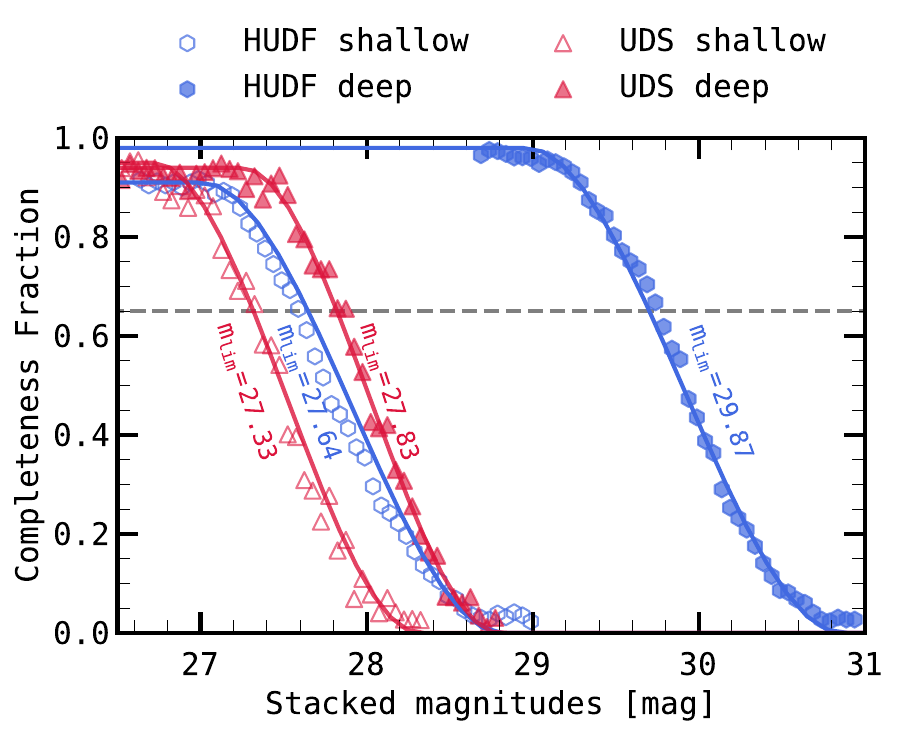}{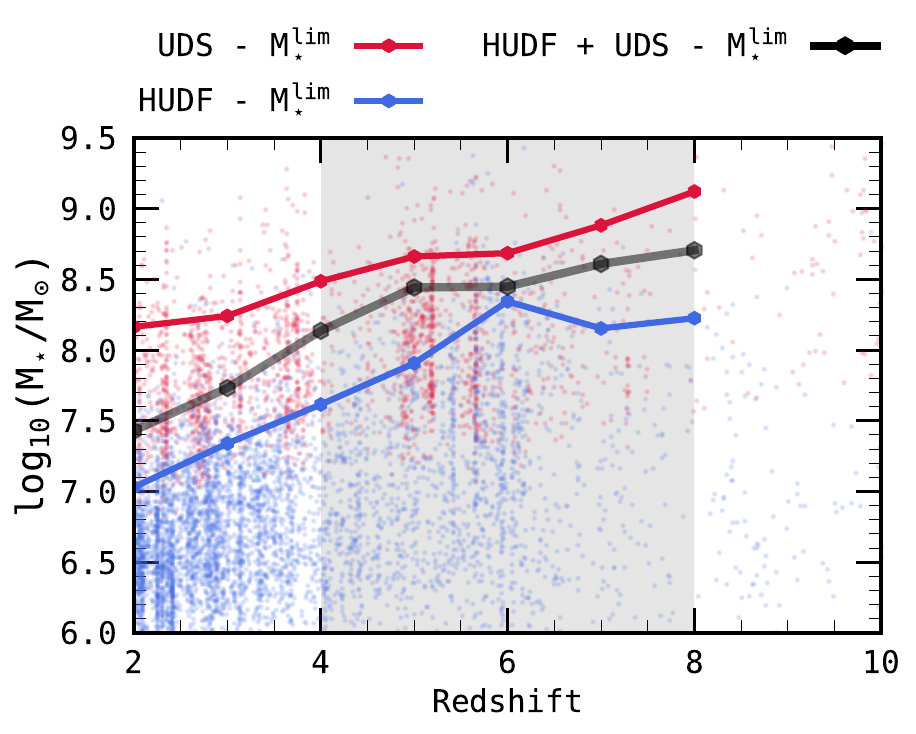}
\caption{\textbf{Left}. Completeness fraction versus stacked magnitude, as obtained from introducing simulated point sources to the detection images. PRIMER-UDS is shown in red and HUDF in blue. Empty symbols represent the shallow part and filled ones the deep part for both fields. The horizontal dashed line shows 65\% completeness. \textbf{Right}. Limiting mass at 90\% completeness over the two fields HUDF and UDS using the method described in \cite{pozzetti_zcosmos_2010}. The value of $\mathcal{M}^{lim}$ is shown as points for all galaxies excluding the brightest 40\% of the sample. The $\mathcal{M}^{lim}(z)$ curve is shown as a line. PRIMER-UDS is shown in red, HUDF in blue and the combination of them in black. The shaded area represents the redshift range $z\simeq4-8$.}
\label{fig:mass_mag_completeness}
\end{figure*}

The 1/Vmax method provides a set of data points to which a functional form can be fitted. We adopt a Schechter function:

\begin{equation}
\label{eq:schechter}
\begin{split}
\Phi(\mathcal{M}|\alpha,M^\ast,\phi^\ast) =  \log(10) \ \phi^\ast e^{-\mathcal{M}/M^\ast} \left( \frac{\mathcal{M}}{M^\ast} \right)^{1-\alpha} \,.
\end{split}
\end{equation}

In order to determine the uncertainties which arise from the 1/Vmax fitting, we have implemented a $\chi^2$ method, by calculating:

\begin{equation}
\chi^2 = \sum \frac{\Phi_{1/Vmax}(\mathcal{M}) - \Phi(\mathcal{M}|\alpha,M^\ast,\phi^\ast)}{\sigma_{1/Vmax}} 
\end{equation}

\noindent between the observed $\Phi_{1/Vmax}(\mathcal{M})$ and fitted $\Phi(\mathcal{M}|\alpha,M^\ast,\phi^\ast)$ data for each stellar-mass bin. Where $\sigma$ is the error (standard deviation) derived for each of the 1/Vmax points. For this purpose, we have developed a parameter space sampler, which provides the best fitting set of parameters $( \alpha,M^\ast,\phi^\ast)$, as well as the $1\sigma$ confidence region for each parameter pair.

After obtaining the GSMF with the 1/Vmax method considering the union of both fields for the calculation (taking into account the effective volume and completeness correction on a galaxy-by-galaxy basis), we fitted the resulting data points with a Schechter function by using a $\chi^2$ minimization method developed for this purpose.

In our calculations, we considered the effects of flux completeness and derived stellar-mass completeness and applied the necessary corrections (Sects. \ref{subsec:method_photometric_completeness} and \ref{subsec:method_mass_completeness}). We also study the influence of Eddington bias \citep{eddington_formula_1913}, which can  potentially affect the shape of the derived GSMF, as detailed in Sect. \ref{subsec:method_eddington}.

\subsection{\rm Photometric completeness} \label{subsec:method_photometric_completeness}
For assessing completeness, we introduced simulated point sources in the detection image. After calculating the fraction of sources recovered for each bin in flux, we have derived a completeness curve (left panel of Fig. \ref{fig:mass_mag_completeness}) which represents the fraction of sources detected versus the total number of sources in the field for each (stacked) magnitude. HUDF (shown in blue) and PRIMER-UDS (shown in red) do not have uniform coverage. Nonetheless, they can be divided into a deep region and shallow region with almost uniform depths. The limiting magnitudes for each band of \textit{JWST} are shown in Table \ref{tab:fields}. The depth of HUDF field is almost uniform for \textit{JWST} data, however this is not the case for \textit{HST} ancillary observations, where the XDF region is significantly deeper with respect to the HUDF (e.g., \citealp{illingworth_hst_2013}).

For the CANDELS complementary data (used to compute the high-mass end of the GSMF), we have obtained the completeness curves for CANDELS-UDS and GOODS-S field from \cite{mortlock_deconstructing_2015}, which follows a similar methodology to ours for computing the photometric completeness.

In this study we have considered only galaxies that lie above 65\% completeness. The number of galaxies with magnitude above the limiting magnitude is $\simeq$1800, conforming the final sample that has been used for the computation of the GSMF. A correction derived from the completeness curves has been introduced when calculating the GSMF as a weight to each galaxy. The limiting stacked image magnitude at 65\% completeness is shown for every region next to the each curve, which can be helpful to asses the difference in depth between the fields used here.

Although faint (low-mass) high-redshift galaxies are expected to be behave like point like sources in the detector, some studies explored the effects of galaxy sizes on the photometric completeness (e.g., \citealt{finkelstein_evolution_2015}). For this reason, we have re-obtained the completeness curve considering extended sources described by a S\'ersic profile \citep{sersic_influence_1963} with index $n=1$ and half-light radius following a uniform distribution between the range of usual values found in \citep[][and references therein]{finkelstein_evolution_2015} (0.4~1.6 kpc). The completeness curve shows a similar limiting magnitude but slightly different shape in the faint and bright ends (at most reaching a $12 \%$ difference in completeness). When re-obtaining the GSMF using the completeness curve derived for extended sources, all 1/Vmax points but the lowest stellar mass point are within the error-bars (the lowest stellar mass 1/Vmax point is the least weighted one in the fitting due to larger uncertainty). In summary, considering that our galaxies are extended instead than point-like does not lead to any significant change in the derived GSMF Schechter parameter values and does not affect any of our conclusions.
\subsection{\rm Stellar mass completeness} \label{subsec:method_mass_completeness}
Estimating the minimum stellar mass at which the sample is complete is crucial for studying the low-mass end of the GSMFs. In the past decades, different methods to estimate the mass completeness have been developed. In particular, for our analysis we follow the method proposed by \cite{pozzetti_zcosmos_2010}, which is commonly used among the literature (e.g., \citealt{davidzon_cosmos2015_2017, weaver_cosmos2020_2022}).

First, galaxies with similar mass-to-light ratios ($\mathcal{M}/L$) to the faintest ones are selected by excluding the 40\% brightest galaxies of the total sample. Their stellar mass $\mathcal{M}^{lim}$ at the limiting flux of the survey $F_{stack}^{lim}$, which in our case is the stacked flux used for the detection (Sect.  \ref{sec:catalogs}), is determined following the relation \citep{weigel_stellar_2016}:
\begin{equation}
\mathcal{M}^{lim} = \mathcal{M} + 0.4 \times (F_{stack}^{lim} - F_{stack}) \,.
\end{equation}
The value of limiting stellar mass at the desired completeness and redshift is defined as the corresponding percentile of the $\mathcal{M}^{lim}(z)$ distribution. The GSMF fit can be considered as secure down to this stellar mass, but below $\mathcal{M}^{lim}(z)$ the mass estimations have non-negligible incompleteness.

Fig. \ref{fig:mass_mag_completeness} shows the $\mathcal{M}^{lim}(z)$ at 90\% completeness curve for PRIMER-UDS (red), HUDF (blue) and for the combination of both (black). The limiting flux of HUDF field is considerably smaller compared to PRIMER-UDS, which results in a lower limiting stellar mass at 90\% completeness. This effect is higher at low redshift, where the difference is $\sim 1 \ \rm{dex}$ and milder at high redshift with $\sim 0.5 \ \rm{dex}$.

In this study we probe values of stellar mass as low as $10^8 {\rm \ M_\odot}$ up to $z\simeq 5$. In other words, we are able to probe down to $\simeq0.5$ dex less massive galaxies than the deepest data of CANDELS \citep{grazian_galaxy_2015} and also down to $\simeq1.75$ dex compared to COSMOS2015 \citep{davidzon_cosmos2015_2017}. We probe down to similar mass ranges as \cite{song_evolution_2016} (that uses very deep \textit{HST} and \textit{SPITZER} data in CANDELS/GOODS-S and HUDF) and \citep{stefanon_galaxy_2021} (that uses very deep SPITZER data in CANDELS/GOODS-N, CANDELS/GOODS-S, COSMOS/UltraVISTA). In literature, there have been studies (e.g., \citealp{kikuchihara_early_2020}) that use gravitational lensing to detect extremely low-mass galaxies, their limiting mass being $\mathcal{M} \simeq 10^7 {\rm \ M_\odot}$ at $z\simeq 6-7$ and $\mathcal{M} \simeq 10^8 {\rm \ M_\odot}$ at $z\simeq 9$.
\section{Results} \label{sec:smf}
By following the method described in Sect. \ref{sec:method}, we parameterize the evolution of the GSMF over cosmic time. First, we have derived the 1/Vmax points with the method described in Sect. \ref{subsec:method_vmax}. Then, we fitted a Schechter Function parametrized as in Eq. \ref{eq:schechter} to the data points by applying a reduced $\chi^2$ fitting method, as described in Sect. \ref{subsec:method_vmax}.

\begin{figure*}[t]
\centering
\includegraphics[width=2.1\columnwidth]{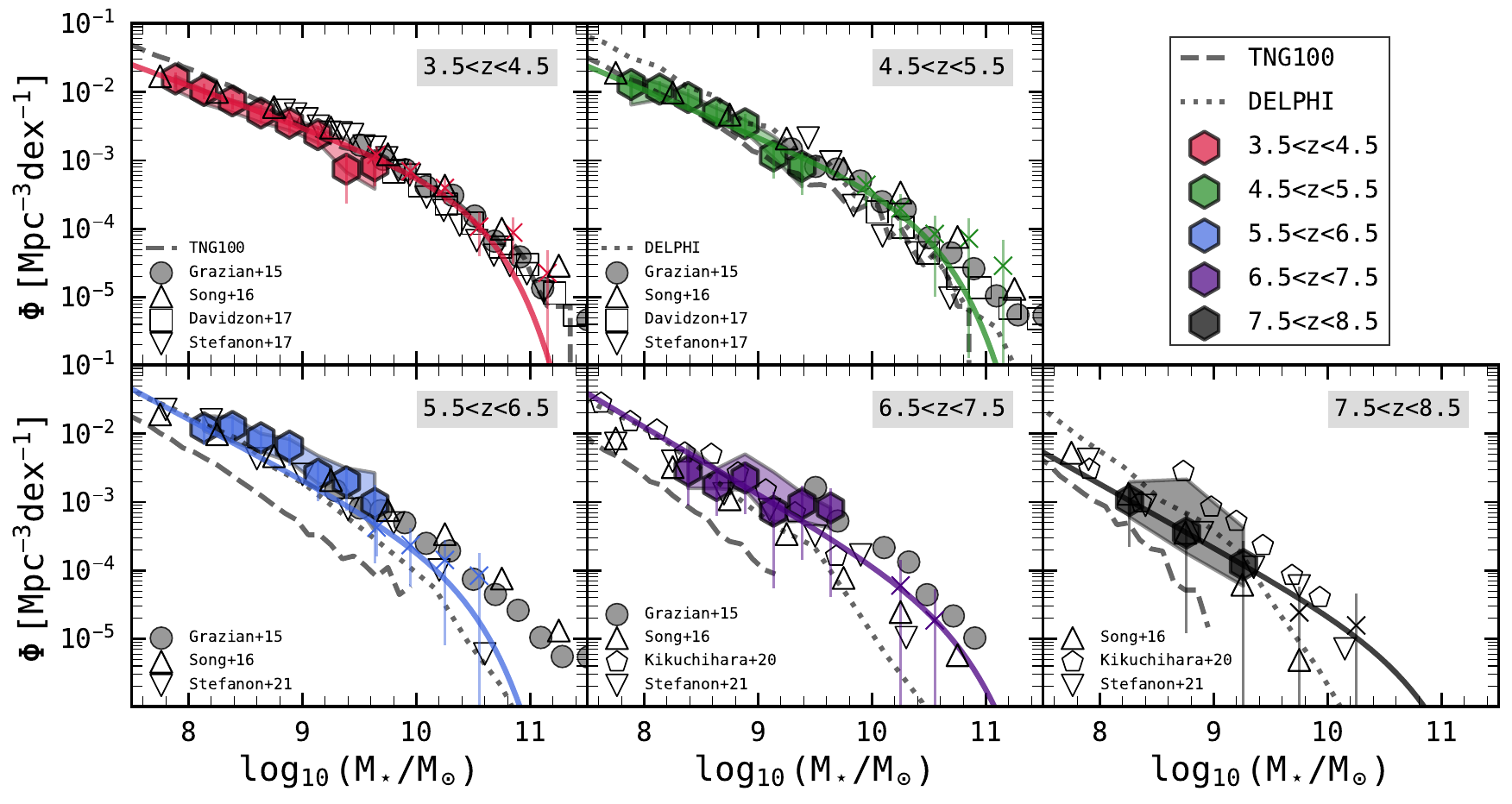}\label{fig:smf_result}
\caption{1/Vmax points (hexagons and crosses for CANDELS ancillary data) and Schechter fits (solid lines) for the GSMF in all redshift bins considered in our study.  The shaded area is the uncertainties obtained using the Monte Carlo method described in Sect. \ref{subsect:photom_uncert}. Errors for the 1/Vmax points include Poisson noise, photometric uncertainty and cosmic variance, as described in Sect. \ref{subsec:method_uncert}. Previous results from the literature are shown as empty or filled polygons: \cite{grazian_galaxy_2015} (filled circles), \cite{song_evolution_2016} (upwards triangles), \cite{davidzon_cosmos2015_2017} (squares), \cite{stefanon_rest-frame_2017} and \cite{stefanon_galaxy_2021} (downwards triangles), \cite{kikuchihara_early_2020} (pentagons). Comparison with simulations is done for Illustris \citep[TNG100,][]{pillepich_first_2018} and \textit{Delphi} \citep{dayal_essential_2014, dayal_alma_2022, mauerhofer_dust_2023}, shown respectively as gray dashed and dotted lines.}
\end{figure*}

However, to account for the biases produced by the propagation of uncertainties to the galaxy number counts (Eddington bias), the Schechter function is first convolved with the stellar mass error distribution (described in  Sect. \ref{subsec:method_eddington}), to obtain the bias-free Schechter parameters. There is no clear consensus on the correct Eddington bias estimation. We refer the reader to \cite{davidzon_cosmos2015_2017} for a thorough discussion on the effects of employing different methods for assessing the Eddington bias. 

\begin{figure}[t]
\centering
\includegraphics[width=0.9\columnwidth]{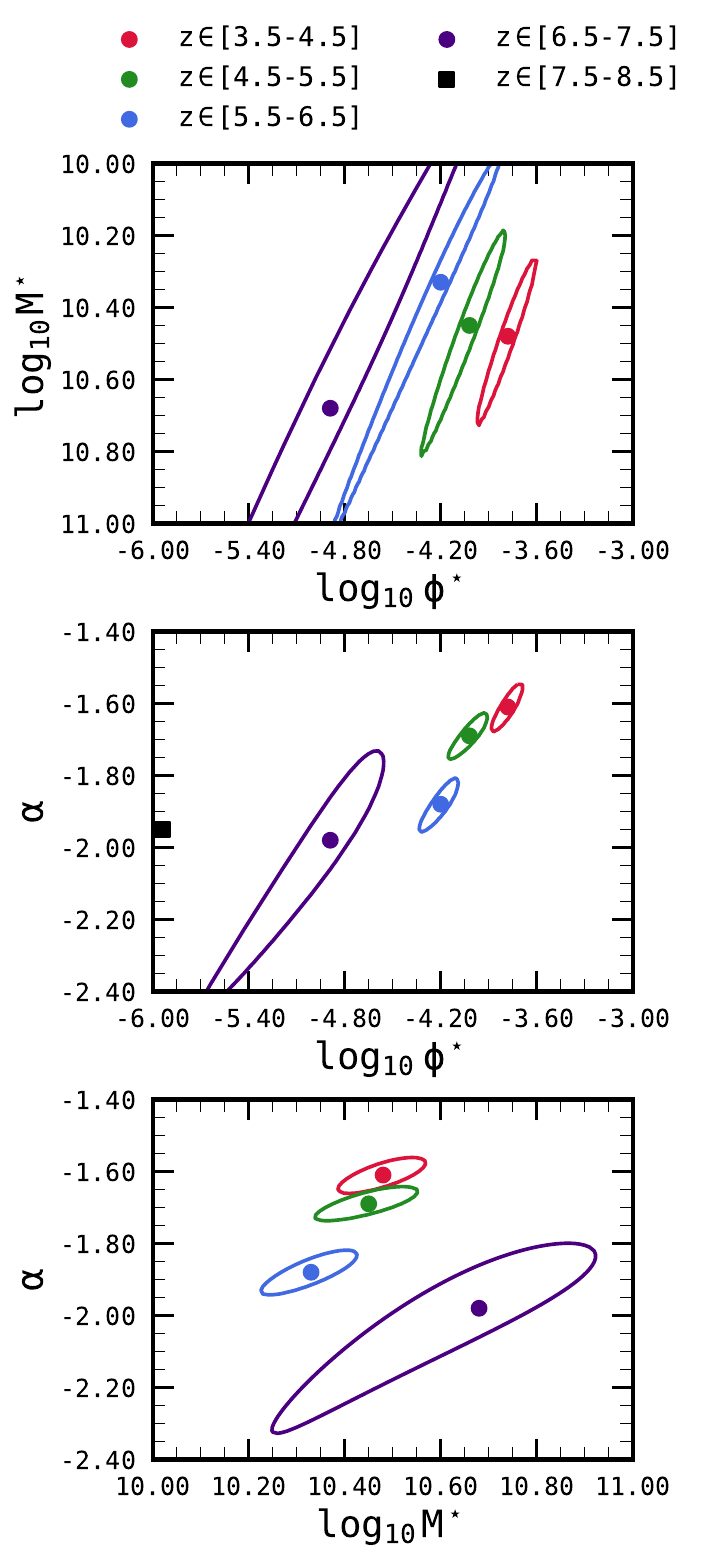}
\caption{Parameter space for the Schechter fits using the $\chi^2$ method described in Section~\ref{subsec:method_vmax}. The point of minimum $\chi^2$ is marked with a color dot, and 1$\sigma$ contours (obtained imposing $\Delta \chi^2_{red} = 1$) are shown with solid color lines. The $M^\ast$ parameter value  has been fixed for the redshift bin z: 7.5 - 8.5 (black square), therefore the corresponding parameter space values are not shown in this figure,  except in the middle panel.}
\label{fig:param_evolution}
\end{figure}

Due to the nature of our dataset, the parameter that is best constrained is the low-mass end slope ($\alpha$). Our \textit{JWST} observations are extremely deep but their area is not large, $\simeq 20$ arcmin$^2$ in total. Characterizing the high-mass end is possible by using ancillary data from larger area surveys (e.g., CANDELS). We will focus our discussion on the evolution of the low stellar-mass end slope and cosmic stellar mass density (CSMD), but we also test for consistency and evolution of the other parameters, $M^\ast$ and $\phi^\ast$. However we refer the reader to \cite{davidzon_cosmos2015_2017} and \cite{weaver_cosmos2020_2022} for a more comprehensive analysis of these other parameters.

We stress that a direct comparison with the literature might result in different absolute values for each Schechter parameter, the low-mass end slope value is coupled to the value of $M^\ast$. This is why a comparison can be made using other quantities such as the (evolution of the) CSMD or the direct 1/Vmax realizations, as shown in Fig. \ref{fig:smf_result}.

\begin{deluxetable}{lcccc}[t]
\tabletypesize{\footnotesize}
\label{tab:smf_results}
\tablecaption{Schechter function best-fit parameters for the GSMF obtained by applying the $\chi^2$ method described in Sect. \ref{subsec:method_vmax} and corrected for Eddington bias (following the method described in Sect. \ref{subsec:method_eddington}). $M^\ast$ for $z\simeq 8$ has been fixed to the value of the previous redshift bin $z\simeq 7$, thus errors are not quoted for this parameter.} 
\tabcolsep=1.5mm
\tablehead{\colhead{Redshift} & $N_{gal}$ & \colhead{$\alpha$} & \colhead{$\log_{10}(M^\ast/{\rm M_\odot})$} & \colhead{$\phi_*$}} 
\startdata
3.5-4.5 & 615 & -1.61$\pm$0.06 & 10.48$\pm$0.15 & (1.65$\pm$0.04)$\times10^{-4}$\\
4.5-5.5 & 467 & -1.69$\pm$0.07 & 10.45$\pm$0.27 & (9.6$\pm$0.9)$\times10^{-5}$\\
5.5-6.5 & 539 & -1.88$\pm$0.09 & 10.33$\pm$0.36 & (6.3$\pm$1)$\times10^{-5}$\\
6.5-7.5 & 147 & -1.98$\pm$0.14 & 10.68$\pm$0.79 & (1.3$\pm$1)$\times10^{-5}$\\
7.5-8.5 & 36 & -1.93$\pm$0.22 & [10.70] & (2.8$\pm$2.1)$\times10^{-6}$\\
\enddata
\end{deluxetable}
\subsection{\rm Evolution of the GSMF with cosmic time} \label{subsec:smf_evol}
The results for the GSMF are shown in Table \ref{tab:smf_results} and Fig. \ref{fig:smf_result}. Hexagons and crosses are 1/Vmax points calculated from \textit{JWST}+\textit{HST} and CANDELS ancillary data, respectively. The continuous line is the $\chi^2$ fit to the 1/Vmax points assuming a Schechter function parametrization, with parameters given in Table \ref{tab:smf_results}. The shaded region is the maximum span of 30 Monte Carlo realizations. For this we recompute GSMF after scrambling the photometry within the error bars and re-obtaining stellar masses and photometric redshifts (see Sect. \ref{subsect:photom_uncert}). This can be understood as an upper limit for the dispersion of the points arising from photometric uncertainties and SED fitting. As shown in \cite{caputi_spitzer_2015}, it can be also used to asses the effect of the propagation of photometric uncertainties to the GSMF (Eddington bias).

We find an evolution of $>1\sigma$ significance for both $\alpha$ and $\phi^\ast$ between each redshift bin. Our results show that the low-mass end slope steepens with cosmic time, following a consistent trend between $z\simeq 7$ and $z\simeq 4$. The value of $\alpha$ changes from close to $\alpha\simeq-2.0$ for $z\simeq 7$ to $\alpha \simeq1.6$ for $z\simeq 4$. This trend holds for the intermediate redshift bins and is discussed in detail in Sect. \ref{sec:smf_alpha_evol}. We find $1\sigma$ agreement in the 1/Vmax points with respect previous studies (e.g., \citealp{duncan_mass_2014, grazian_galaxy_2015, song_evolution_2016, stefanon_galaxy_2021}).

A number of studies, from \cite{perezgonzalez_stellar_2008} to  more recent ones like \cite{stefanon_galaxy_2021}, find that the normalization factor $\phi^\ast$ evolves with cosmic time following the cosmic stellar mass buildup. \cite{grazian_galaxy_2015} find the that $\phi^\ast_{z\simeq 4}/\phi^\ast_{z\simeq 7} \simeq 20^{+3000}_{-19}$. \cite{song_evolution_2016} find  $\phi^\ast_{z\simeq 4}/\phi^\ast_{z\simeq 7} \simeq 48^{+268}_{-41}$. According to our Schechter fits $\phi^\ast$ increases by a factor of $130^{+210}_{-50}$ from $z\simeq 7$ to $z\simeq 4$, compatible within the previous measurements within the uncertainties.

For $M^\ast$, values are compatible with each other within $1\sigma$. We emphasize that probing $M^\ast$ is not the main goal of this study, and that the high-mass is taken from ancillary CANDELS data. Recent studies that probe the high-mass end of the GSMF conclude that there is not a clear evidence on the evolution of $M^\ast$ (e.g., \citealp{davidzon_cosmos2015_2017}), however evidences for such a evolution have been found for lower redshift samples (e.g., \citealp{adams_evolution_2021}). We find that $M^\ast$ values are, in general, consistent with the ones derived by \cite{davidzon_cosmos2015_2017} and \cite{grazian_galaxy_2015}. However, some differences are expected to arise from the different methodology used for Eddington bias correction.

Compared to more recent studies, the resulting values of the GSMF best fit parameters are compatible to the ones derived by \cite{song_evolution_2016}, that also made use of CANDELS and HUDF fields for their study. \cite{stefanon_galaxy_2021} sees a non-steepening $\alpha$ and consistently lower values of $M^\ast$, although we stress that our constraint in $M^\ast$ is not strong because of the small area covered by our dataset and comes mainly from existing CANDELS data. We thoroughly study the evolution of $\alpha$ in Sect. \ref{sec:smf_alpha_evol}.

\begin{deluxetable}{l|ccccc}[h!]
\tabletypesize{\footnotesize}
\label{tab:smf_values}
\tablecaption{The GSMF computed with the 1/Vmax method. The highest redshift bin $z\simeq 8$ is twice as large as the other redshift bins, in order to increase the statistics and, thus, robustness of our result. The points that lie under the stellar mass completeness limits or that have $\leq1$ galaxy counts are not quoted in this Table.} 
\tabcolsep=1.5mm
\tablehead{$\log_{10}(\mathcal{M}/{\rm M_\odot})$ & \multicolumn{5}{c}{$\phi \, [\rm{Mpc^{-3}dex^{-1}}]\, \times 10^4$}  \\ 
 & \colhead{$z\simeq 4$} & \colhead{$z\simeq 5$} & \colhead{$z\simeq 6$} & \colhead{$z\simeq 7$} & \colhead{$z\simeq 8$}}
\startdata
7.75-8.00 & 155$\pm$39 & 128$\pm$49 & -          & -         & -         \\ \cline{1-6}
8.00-8.25 & 105$\pm$27 & 110$\pm$36 & 121$\pm$53 & -         & \multirow{2}{*}{11$\pm$8}  \\
8.25-8.50 & 74$\pm$19  & 83$\pm$27  & 127$\pm$55 & 29$\pm$20 &           \\ \cline{1-6}
8.50-8.75 & 49$\pm$14  & 49$\pm$18  & 87$\pm$36  & 18$\pm$11 & \multirow{2}{*}{4$\pm$3}   \\
8.75-9.00 & 35$\pm$11  & 34$\pm$13  & 65$\pm$32  & 22$\pm$16 &           \\ \cline{1-6}
9.00-9.25 & 24$\pm$8   & 12$\pm$6   & 25$\pm$15  & 7$\pm$7   & \multirow{2}{*}{1$\pm$1}   \\ 
9.25-9.50 & 7$\pm$5    & 8$\pm$5    & 20$\pm$12  & 9$\pm$8   &           \\\cline{1-6}
9.50-9.75 & 8$\pm$4    & -          & 10$\pm$8   & 8$\pm$8   & -         \\
\enddata
\end{deluxetable}

As seen in Fig. \ref{fig:smf_result}, we probe the GSMF until $z\simeq8.5$. Due to the few number of galaxies in each stellar mass bin, for both \textit{JWST} and CANDELS data, the value of $M^\ast$ has been constrained between the range of values found for the other redshift bins in the last redshift bin $z\simeq7.5-8.5$. This is why the value of $\alpha$ derived for this redshift bin ($z\simeq8$) is only given as a tentative value, and should be investigated further in future observations.

In Fig. \ref{fig:smf_result} we compare our results to the GSMF computed from IllustrisTNG \citep[][TNG100]{pillepich_first_2018} as a grey dashed line. The agreement is good for $z\simeq4$ and $z\simeq5$. Nonetheless, the low stellar-mass end has a steeper slope for the TNG100 determination at $z\simeq4$. The low-mass end slopes of TNG100 are similar to the ones from our analysis for all redshift bins, but for $z\simeq6$ and for $z\simeq7$, the normalization of the GSMF is significantly different. TNG100 predicts a lower density of galaxies at all stellar masses. The low-mass end only behaves differently at $z\simeq8$, where our observations also suffer from larger uncertainties.

We also include the GSMF obtained from \textit{Delphi} (Dark Matter and the emergence of galaxies in the epoch of reionization; \citealp{dayal_essential_2014, dayal_alma_2022, mauerhofer_dust_2023}). The overall agreement of \textit{Delphi} with our realization of the GSMF is better than the one of TNG100 for $z\simeq 6$, $z\simeq 7$. This is not true for the lower and higher redshift bins, where we find disagreement with \textit{Delphi} for both high and low-mass end, specially for $z\simeq 8$. From the lack of data in the high-mass end, TNG100 volume is not large enough for capturing the galaxies that conform the massive end of the GSMF, specially above $\mathcal{M} > 10^9 {\rm \ M_\odot}$ and $z\gtrsim 6$.

In Fig. \ref{fig:param_evolution}, we show the parameter space for Fig. \ref{fig:smf_result} fits of the 1/Vmax points. It depicts the best-fit value as solid points and the $1\sigma$ confidence level as continuous contours. The $1\sigma$ confidence level has been determined by imposing for $\Delta \chi^2_{red} = 1$. This shows that the evolution of $\alpha$ and $\phi^\ast$ is significant within contiguous redshift bins, and that there exists a steepening of $\alpha$ between $z\simeq4$ and $z\simeq7$. It can also be seen that there is statistically significant evolution of $\phi^\ast$ to higher values toward lower redshifts.

We performed a careful analysis of the sources of uncertainty that may affect the GSMF. A detailed explanation is given in Appendix \ref{sect:gsmf_uncert}. We conclude that these errors have a minor impact on our results and do not siginficantly change any of our conclusions. We include the best fit Schechter parameters not accounting for the Eddington bias in Table \ref{tab:smf_results_noncorr}. 

\begin{deluxetable}{lcccc}[h!]
\tabletypesize{\footnotesize}
\label{tab:smf_results_noncorr}
\tablecaption{Schechter function best-fit parameters for the GSMF obtained using the $\chi^2$ method described in Sect. \ref{subsec:method_vmax} without Eddington bias correction. $M^\ast$ for $z\simeq 8$ has been fixed to the value of the previous redshift bin $z\simeq 7$, thus errors are not quoted for this parameter.} 
\tabcolsep=1.5mm
\tablehead{\colhead{Redshift} & $N_{gal}$ & \colhead{$\alpha$} & \colhead{$\log_{10}(M^\ast/{\rm M_\odot})$} & \colhead{$\phi_*$}} 
\startdata
3.5-4.5 & 615 & -1.53$\pm$0.05 & 10.73$\pm$0.16 & (1.1$\pm$0.3)$\times10^{-4}$\\
4.5-5.5 & 467 & -1.87$\pm$0.06 & 10.79$\pm$0.06 & (2.8$\pm$0.4)$\times10^{-5}$\\
5.5-6.5 & 539 & -2.03$\pm$0.08 & 10.43$\pm$0.19 & (5$\pm$2)$\times10^{-5}$\\
6.5-7.5 & 147 & -1.89$\pm$0.14 & 10.74$\pm$0.42 & (1.1$\pm$0.9)$\times10^{-5}$\\
7.5-8.5 & 36 & -1.85$\pm$0.25 & [10.70] & (4$\pm$2)$\times10^{-6}$\\
\enddata
\end{deluxetable}

\section{\rm Evolution of the low-mass end slope with cosmic time}
\label{sec:smf_alpha_evol}

A key goal of this work is to constrain the GSMF low-mass end slope at high $z$, so here we analyse its evolution with redshift.
 While a direct comparison of $\alpha$ can be affected by numerous systematic effects such as Eddington Bias estimation, it can be helpful to visualize the evolution of this parameter with cosmic time. And although $\alpha$ is coupled to the other Schechter parameters (mainly $M^\ast$), we find a reasonable agreement with the literature for the $M^\ast$ values.
 
\begin{figure}[t]
\centering
\includegraphics[width=1\columnwidth]{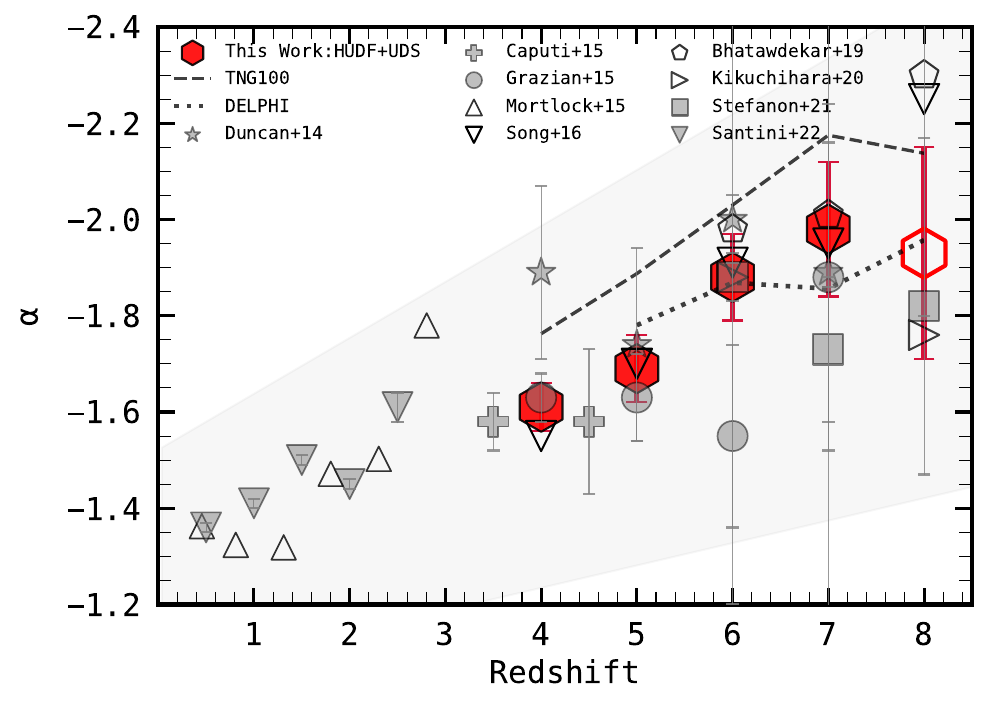}
\caption{Evolution of the faint-end slope as a function of redshift. Points from our study are shown as red hexagons. Errors have been estimated from the $\chi^2$ fit by taking into account all contributions described in Sect. \ref{fig:smf_uncert}. Points from previous studies have been included as black empty or gray shaded symbols: \cite{gonzalez_evolution_2011}, \cite{duncan_mass_2014}, \cite{caputi_spitzer_2015}, \cite{grazian_galaxy_2015}, \cite{mortlock_deconstructing_2015}, \cite{song_evolution_2016}, \cite{davidzon_cosmos2015_2017}, \cite{stefanon_galaxy_2021}, \cite{santini_stellar_2022}. Predictions from Illustris TNG100 and \textit{Delphi} are shown as dashed and dotted grey lines, respectively.}
\label{fig:alpha_evolution}
\end{figure}

 Fig. \ref{fig:alpha_evolution} shows the comparison between the value of $\alpha$ obtained in our analysis and the values fromliterature. It is clear that $\alpha$ is becoming steeper towards high redshifts. This implies that the evolution of the GSMF is not driven by a pure (number) density evolution.
 
The agreement with \cite{song_evolution_2016} is remarkably good for all redshift bins, except for the last one ($z\simeq8$), where $\alpha=-2.25\pm0.5$ for \cite{song_evolution_2016} and $\alpha=-1.93\pm0.22$ for our analysis. Our results also agree with the ones from \cite{stefanon_galaxy_2021}, where the biggest difference between our and their results occurs at $z\simeq 7$ where $\alpha=-1.73\pm0.15$ for \cite{stefanon_galaxy_2021} and $\alpha=-1.98\pm0.14$ for our analysis. However, these values are still compatible within $1\sigma$ in both cases.

We find a good agreement with \cite{grazian_galaxy_2015} for all redshift bins except for $z\simeq 6$, where our value of $\alpha$ is significantly higher: $\alpha=-1.55\pm0.19$ for \cite{grazian_galaxy_2015} and $\alpha=-1.88\pm0.09$ for our analysis. Note that the \cite{grazian_galaxy_2015} limiting mass is $\mathcal{M}^{lim} \simeq 10^{9.2} {\rm \ M_\odot}$ while ours is $\mathcal{M}^{lim} \simeq 10^{8.4} {\rm \ M_\odot}$, i.e. our constraint on the low-mass end slope $\alpha$ is more robust.

\cite{davidzon_cosmos2015_2017} find an overall higher value of $\alpha$ with respect to our analysis but also with respect to \cite{grazian_galaxy_2015} and the rest of literature. This can be due to the large area of the COSMOS field \citep{scoville_cosmic_2007} but shallower data. The limiting stellar mass for $z\simeq 5$ is $\mathcal{M}^{lim} \simeq 10^{10} {\rm \ M_\odot}$. The effect of the Eddington Bias estimation can also play a role in their estimations, as discussed in \cite{davidzon_cosmos2015_2017}. The $\alpha$ values obtained by \cite{davidzon_cosmos2015_2017} are not included in the comparison in Fig. \ref{fig:alpha_evolution}.

From our analysis it follows that the low-mass end slope steepens with cosmic time within $z\simeq 4-7$, from $\alpha = -1.61\pm0.05$ for $z\simeq 4$ to $\alpha=-1.98\pm0.14$ for $z\simeq 7$. We find a good agreement with recent studies that probe the low-mass end of the GSMF (e.g., \citealp{song_evolution_2016, bhatawdekar_evolution_2019, stefanon_galaxy_2021}) in the steepening of $\alpha$ with redshift. We cannot confirm or rule out the presence of a turnover above $z\simeq 8$. Altough the 1/Vmax points for $z\simeq 8$ agree with other studies, we find different Schechter parameters. We consider the value of $\alpha$ at $z\simeq 8$ obtained from our analysis as tentative (we include it as empty symbols in Figs. \ref{fig:alpha_evolution} and \ref{fig:smd}).

However, in \cite{kikuchihara_early_2020} there is no significant evolution of $\alpha$ between $z\simeq 6-7$ and $z\simeq 9$ for their study of the GSMF in the Hubble Frontier fields \citep[HFF,][]{lotz_frontier_2017}. \cite{bhatawdekar_evolution_2019} do find a steepening of $\alpha$ (within the HFF cluster MACSJ0416.1-2403) between $z\simeq 6$ and $z\simeq 9$, this is, identical redshift range studied by \cite{kikuchihara_early_2020}. Both of them probe a similar range of stellar masses, reaching $\mathcal{M} \simeq 10^8 {\rm \ M_\odot}$ for $z\simeq 8$, although arguably \cite{kikuchihara_early_2020} uses a larger dataset.

Finally, both TNG100 and \textit{Delphi} predict a steepening low-mass end slope with cosmic time within $z\simeq 4$ ($z\simeq 5$ for \textit{Delphi}) and $z\simeq 8$. The values from \textit{Delphi} are located within the error bars of our measurements for all redshifts. In the case of TNG100, we observe a rigid shift towards steeper values of $\alpha$ that holds roughly constant for all redshift bins analyzed herein.
\section{EVOLUTION OF THE COSMIC STELLAR MASS DENSITY} \label{sec:stellar_mass_density}
\begin{figure}[t]
\centering
\includegraphics[width=1\columnwidth]{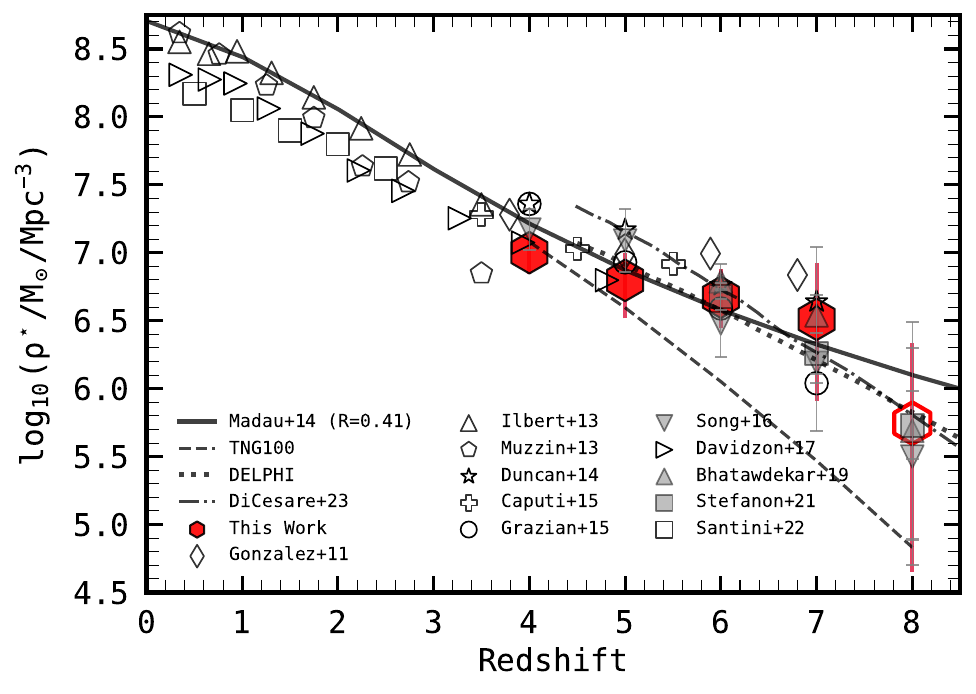}
\caption{Evolution of the cosmic stellar mass density as a function of redshift. We include our own realization (red hexagons) and literature (\cite{gonzalez_evolution_2011}, \cite{ilbert_mass_2013}, \cite{muzzin_evolution_2013}, \cite{duncan_mass_2014}, \cite{caputi_spitzer_2015}, \cite{grazian_galaxy_2015}, \cite{song_evolution_2016}, \cite{davidzon_cosmos2015_2017}, \cite{bhatawdekar_evolution_2019}, \cite{stefanon_galaxy_2021}, \cite{santini_stellar_2022}). We include the CSMD resulting from integrating the cosmic star formation history of \cite{madau_cosmic_2014} with a recycled fraction of 41\% as a black solid line. We also compare our results with the predictions of theoretical models, namely \textit{Delphi} (dotted black line), Illustris TNG100 (dashed black line) and \cite{dicesare_assembly_2023} (dash-dotted black line).}
\label{fig:smd}
\end{figure}
The cosmic stellar mass density (CSMD, $\rho_\star$) accounts for the total mass in form of stars that has been formed until a certain redshift or cosmic time. The CSMD provides a more global overview of the stellar mass buildup compared to the GSMF. We can derive the CSMD from the GSMF computed for a given redshift bin (with mean redshift $z_i$) by integrating the GSMF ($\Phi^{z_i}$) for reach redshift bin ($z_i$) over a range of stellar masses ($\mathcal{M}_{inf}$, $\mathcal{M}_{sup}$) \citep{fontana_galaxy_2006}:
\begin{equation}
\rho_\star^{z_i} = \int^{\mathcal{M}_{inf}}_{\mathcal{M}_{sup}}\Phi^{z_i}(\mathcal{M}) \times \mathcal{M} \, d\mathcal{M} \,.
\end{equation}
In the literature, the values of stellar masses to compute the integral are usually chosen as $\mathcal{M}_{inf} = 10^8 {\rm \ M_\odot}$ and $\mathcal{M}_{sup} = 10^{13} {\rm \ M_\odot}$ (e.g., \citealp{elsner_impact_2008, stark_keck_2013, kikuchihara_early_2020}). This is the convention followed in this work as well. Values for the CSMD are shown in Table \ref{tab:smd_results}. We used the GSMF best-fit parametrizations shown in Table \ref{tab:smf_results} after correction for Eddington bias for calculating the CSMD.

Fig. \ref{fig:smd} shows the evolution of the CSMD from $z\simeq 3.5$ to $z\simeq 8.5$. We find an overall good agreement with previous studies. In the common redshifts, $z\simeq 4$ and $z\simeq 5$, our values are very similar to the ones of \cite{davidzon_cosmos2015_2017}.

\cite{song_evolution_2016} cover the same redshift range as us, finding compatible results for all of them within the uncertainties, although we find an overall flatter evolution of the CSMD.

Our results on the CSMD are very similar to those by \cite{bhatawdekar_evolution_2019} for redshifts $z\simeq 6$, $z\simeq 7$ and $z\simeq 8$. Nevertheless, the values of the CSMD are more uncertain at higher redshifts. This is also the case for \cite{stefanon_galaxy_2021}, except for $z\simeq 7$ were their CSMD is more similar to the one by \cite{song_evolution_2016}, but compatible with our estimation within $1\sigma$.

All the CSMD values show the same behavior at $z\simeq 8$, but in our case incompleteness can be affecting the results. To compute the CSMD down to $\mathcal{M}_{inf} = 10^8 {\rm \ M_\odot}$ we had to extrapolate below the mass completeness limit. Also, the value of $M^\ast$ was fixed for the best-fit Schechter function. For all the reasons above, our estimation of the GSMF and CSMD at $z\simeq 8$ has to be taken as a tentative one.

We have computed the CSMD integrating the cosmic star formation history by \cite{madau_cosmic_2014}, considering a recycled fraction of $R=0.41$ ($41\%$), compatible with the Chabrier IMF \cite{chabrier_galactic_2003}. We can reproduce the flattening of the CSMD, but our point at $z\simeq 8$ falls below the prediction of \cite{madau_cosmic_2014}. However, only very few measurements were available to constrain the CSMD at $z\simeq 8$ at the time of the \cite{madau_cosmic_2014} analysis.

Fig. \ref{fig:smd} also shows the CSMD calculation for TNG100 simulation as a black dashed line, integrating over the GSMF presented in Fig. \ref{fig:smf_result}. The agreement is good for redshifts $z\simeq 4$ and $z\simeq 5$. From $z\simeq 6$, however, the CSMD of TNG100 decreases rapidly meanwhile the our results and the ones from literature suggest that the shape of the CSMD flattens with higher redshifts. 

We have also included the recent results by \cite{dicesare_assembly_2023} and \textit{Delphi} in the comparison, as dotted and dashed-dotted black lines. The CSMD of \cite{dicesare_assembly_2023} is significantly higher than the one of TNG100 at all redshifts, and the agreement with our results is better at high redshifts ($z\simeq 6$, $z\simeq 7$ and $z\simeq 8$). \cite{dicesare_assembly_2023} make use of the hydrodynamical code \textsc{DustyGadget} \citep{graziani_assembly_2020}. It accounts for the dust produced by stellar populations and results in a better agreement with the observed CSMD obtained in this work. The CSMD predicted by \textit{Delphi} is in good agreement with the predictions of \citep{dicesare_assembly_2023} for redshifts $z\gtrsim7$. For lower redshifts, \textit{Delphi} behaves more similarly to the best-fit line produced by \citealp{madau_cosmic_2014} and our own data points.
\section{Summary and Conclusions} \label{sec:conclusion}
\begin{deluxetable}{cc}[t]
\tabcolsep=8mm
\tabletypesize{\footnotesize}
\centering
\label{tab:smd_results}
\tablecaption{Results for the CSMD obtained from our best fit Schechter functions, integrating from $\mathcal{M} = 10^{8} {\rm M_\odot}$ to $\mathcal{M} = 10^{13} {\rm \ M_\odot}$.} 
\tablehead{\colhead{Redshift} & \colhead{$\log_{10}(\rho_*/{\rm \ M_\odot}/\rm{Mpc^{-3}})$}} 
\startdata
3.5-4.5 & 7.00$^{+0.14}_{-0.16}$\\
4.5-5.5 & 6.79$^{+0.20}_{-0.28}$\\
5.5-6.5 & 6.67$^{+0.21}_{-0.23}$\\
6.5-7.5 & 6.51$^{+0.42}_{-0.60}$\\
7.5-8.5 & 5.75$^{+0.59}_{-1.10}$\\
\enddata
\end{deluxetable}

In this paper we study the  GSMF and its evolution with redshift between $z\simeq 4$ and $z\simeq 8$. By making use of new observations from \textit{JWST} up to $\sim 5 \, \rm \mu m$ in the PRIMER-UDS and HUDF fields, we are able to probe individual galaxies down to stellar masses  $\mathcal{M} \sim 10^{8} {\rm \ M_\odot}$ up to $z\simeq 8$ for the first time. Until now, this low stellar-mass regime at high redshifts was only  reached using lensing clusters (e.g., \citealp{bhatawdekar_evolution_2019, rinaldi_galaxy_2022}), or statistically with stacking analysis (e.g., \citealp{song_evolution_2016}), and in most cases the stellar-mass determinations were based on long spectral extrapolations. In contrast, the stellar masses derived here are based on the directly measured rest-frame optical light of galaxies up to the highest redshifts.

To compute the GSMF, we adopt  a technique based on the 1/Vmax method, using a $\chi^2$ minimization method that samples the parameter space and provides us with both  best-fit values and uncertainties  for the GSMF Schechter-function parameters. Before doing so, we assess all the systematics and sources of uncertainty that can affect the GSMF, such as the propagation of photometric uncertainties to the photometric redshifts and stellar masses (the so-called Eddington bias) and cosmic variance.

To account for the Eddington Bias, we first derive the error distributions on the stellar mass for each redshift and stellar mass bin. We choose a combination of Gaussian and Student-T kernels, allowing for skewness in the final distribution. We then convolve the pure Schechter function with the error distributions to correct for the effects of Eddington bias. This affects the shape of the GSMF mainly in the high and low-mass ends. In particular, the value of the low-mass end slope becomes less steep after accounting for the Eddington bias, and $M^\ast$ moves towards lower masses.


The GSMF obtained here is in good agreement with the literature based on pre-\textit{JWST} datasets (e.g., \citealp{grazian_galaxy_2015, song_evolution_2016, bhatawdekar_evolution_2019}). This overall good agreement could be considered to some extent fortuitous, as our study is one of the first one to directly probe down to the typical stellar masses of (local) satellite galaxies at high redshifts with significant level of statistics. 

Also, in agreement with some recent results (albeit not all), we find a significant ($>1\sigma$) evolution of the faint-end slope $\alpha$ with cosmic time, corresponding to a steepening of the low-mass end of the GSMF toward earlier times. Instead, we do not see any significant evolution in $M^\ast$.

When comparing to TNG100, we find that the agreement is good for the lowest redshifts analysed here ($z\simeq4-5$). The faint-end slope is in broad agreement up to higher redshifts, but the normalization  of TNG100 falls significantly below ours for at $z\gsim5$, which suggests that these models systematically underestimate the overall number density of galaxies at high redshifts. With respect to \textit{Delphi} we find a overall better agreement specially in the higher redshifts $z\gtrsim 5$, in both the GSMF and the CSMD. 

The number density of galaxies with masses between $\log_{10}(\mathcal{M}/{\rm \ M_\odot}) = 8.25-8.75$ (the lowest mass regime within our completeness in estellar mass) grows by a factor of 1.5 between redshifts $z\simeq 4$ and $z\simeq 7$ according to our realization of the GSMF, meanwhile \textit{Delphi} predicts a increase factor of 2.8. However, the number density of these galaxies in \textit{Delphi} is systematically higher at the low redshifts $z\simeq4.5$.

Finally, we compute the cosmic stellar mass density (CSMD) and study its evolution. We find a broad agreement with the recent literature, but at the same time predict a tentative shallower evolution from $z\simeq 4$ to $z\simeq 7$. Unfortunately, the large errors for the CSMD data point at $z\simeq8$ do not allow us to confirm whether the flattening trend continues up to such high redshift. TNG100 shows a similar behaviour to our results for at $z\simeq 4-5$, however their CSMD falls rapidly below ours (and most literature) for $z \gtrsim 6$. New models such as those by \cite{dicesare_assembly_2023} seem to reproduce better the evolution of the CSMD for these higher redshifts.
\section{ACKNOWLEDGMENTS} \label{sec:ACKNOWLEDGMENTS}
The authors would like to thank P. C\'aceres-Burgos and J. M\'endez-Gallego for useful scientific discussions; and to Pratika Dayal and Valentin Mauerhofer for providing numerical predictions of the DELPHI galaxy formation model.

RN, VK and KIC acknowledge funding from the Dutch Research Council (NWO) through the award of the Vici Grant VI.C.212.036. This work is based on observations made with the NASA/ESA/CSA James Webb Space Telescope. The data were obtained from the Mikulski Archive for Space Telescopes at the Space Telescope Science Institute, which is operated by the Association of Universities for Research in Astronomy, Inc., under NASA contract NAS 5-03127 for \textit{JWST}. These observations are associated with \textit{JWST} programs GO \#1963, GO \#1895, GO \#1837. The authors acknowledge the team led by coPIs C. Williams, M. Maseda and S. Tacchella, and PI P. Oesch, for developing their respective observing programs with a zero-exclusive-access period. Also based on observations made with the NASA/ESA Hubble Space Telescope obtained from the Space Telescope Science Institute, which is operated by the Association of Universities for Research in Astronomy, Inc., under NASA contract NAS 526555.

\textit{Software}: \texttt{AstroPy} \citep{collaboration_astropy_2022}, \texttt{dustmaps} \citep{green_dustmaps_2018}, \textsc{LePHARE} \citep{arnouts_lephare_2011}, \texttt{Matplotlib} \citep{hunter_matplotlib_2007}, \texttt{NumPy} \citep{harris_array_2020}, \textsc{Photutils} \citep{bradley_astropyphotutils_2022}, \textsc{Python} \citep{van_rossum_python_1995}, \texttt{SciPy} \citep{virtanen_scipy_2020}, \textsc{Source Extractor} \citep{bertin_sextractor_1996}, \textsc{TOPCAT} \citep{taylor_topcat_2017}, \textsc{WebbPSF} \citep{perrin_updated_2014}.

\textit{Facilities}: \textit{HST}, \textit{JWST}

\newpage
\appendix

\section{\rm Uncertainties in the determination of the GSMF}\label{sect:gsmf_uncert}
\subsection{Poisson uncertainty}\label{subsect:poisson_uncert}
Poisson statistics describe random processes with small number counts, which is the case of the 1/Vmax method used here to compute the GSMF for our galaxy samples. The error associated with having $N$ counts in a stellar mass and redshift bin is $\sqrt{N}$.

For bins with no detections at all, upper and lower limits can be obtained from the Poisson distribution, as described in \cite{gehrels_confidence_1986} and in \cite{ebeling_improved_2003} (where a more robust estimation of the upper limits is provided). However, we have opted not to include them in our calculation, as very few bins have no detections in both our data and the ancillary CANDELS dataset.

\begin{figure*}[ht]
\centering
\includegraphics[width=0.65\columnwidth]{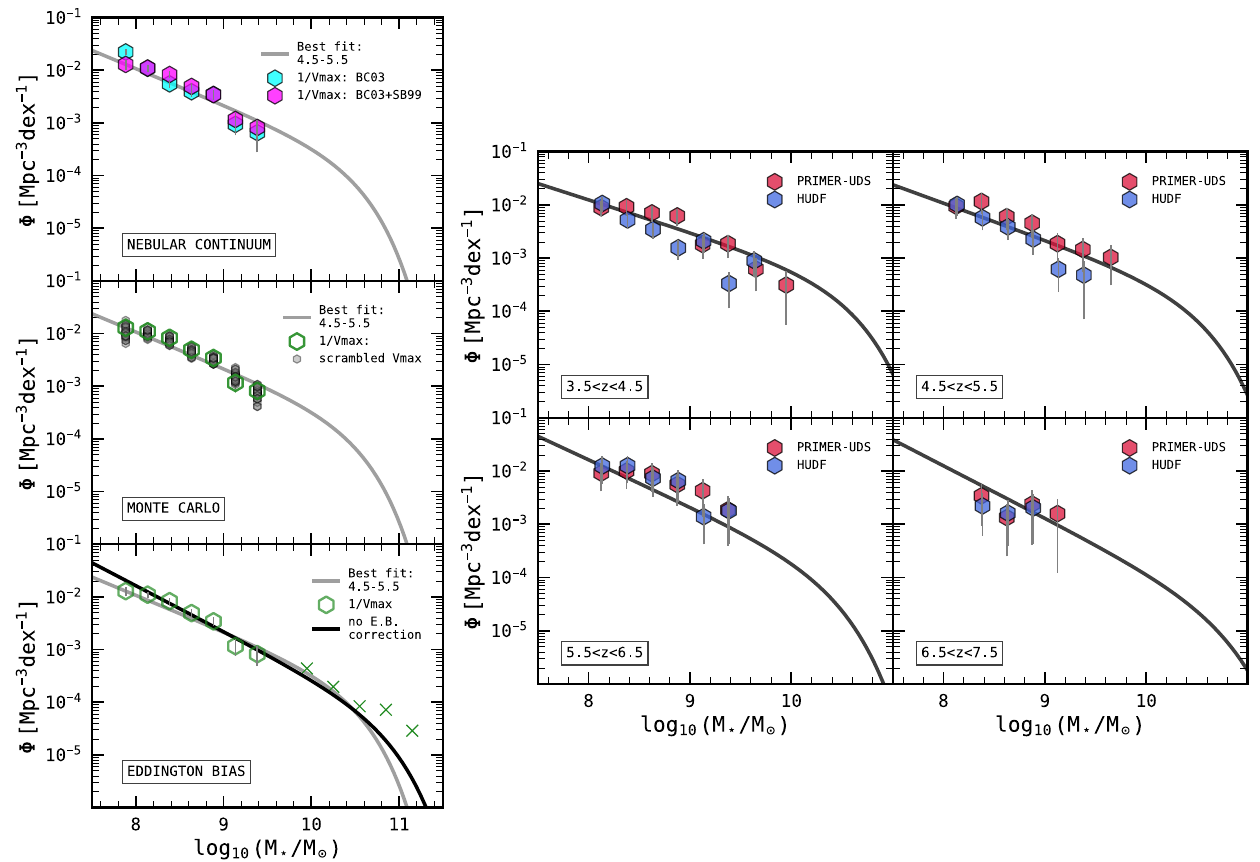}
\caption{\textbf{Left}: the bottom panel shows the effect of Eddington Bias correction on the Schechter fit for the GSMF, note the effect on both the high and low-mass end by making the GSMF less steep and reducing the value of $M^\ast$. The middle panel  shows the 30 realizations of the GSMF after randomizing the photometry within the errors, where the scatter is higher for the lower and highest mass points. The upper panel shows the effect of accounting for nebular emission, all the 1/Vmax points are in good agreement within 1$\sigma$ errors. The lowest mass point showing the biggest difference. \textbf{Right}: the effect of cosmic variance for redshift bins $z\simeq 4,5,6,7$, by plotting the 1/Vmax for PRIMER-UDS (red) and HUDF (blue) separately.}
\label{fig:smf_uncert}
\end{figure*}
\label{subsec:method_uncert}
\begin{figure*}[t]
\centering
\includegraphics[width=0.7\columnwidth]{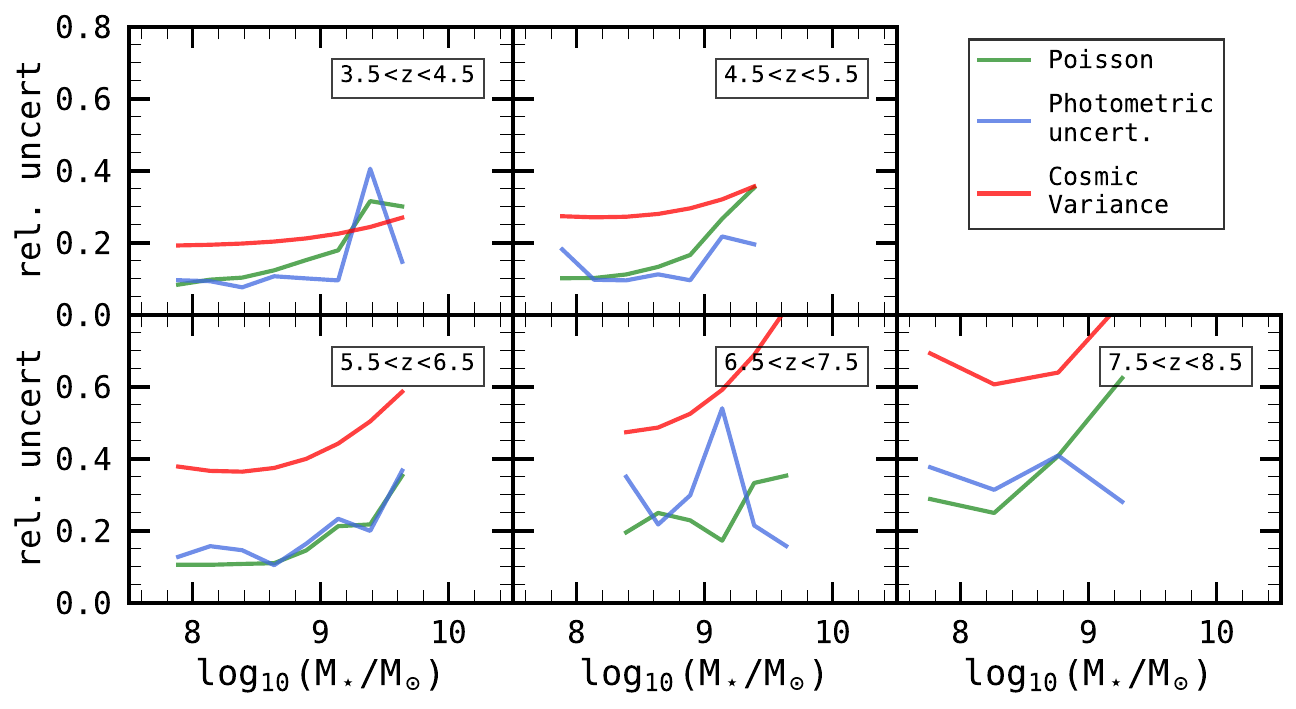}
\caption{Error budget for all redshift bins and stellar masses covered by \textit{JWST}. Poisson uncertainty (Sect. \ref{subsect:poisson_uncert}) is shown as a green continuous line, photometric uncertainty (Sect. \ref{subsect:photom_uncert}) is shown as a continuous blue line and cosmic variance (Sect. \ref{subsect:cosmic_variance}) is shown as a continuous red line.}
\label{fig:err_budget}
\end{figure*}
\subsection{Photometric uncertainty}\label{subsect:photom_uncert}
Photometric uncertainty is the one of the main random uncertainties contributors to the error budget. Errors in photometry propagate into the stellar mass and photometric redshift obtained from the SED fitting. This errors are especially relevant for faint galaxies, and their correct assessment is crucial for an appropriate description of the GSMF low-mass end.

In order to estimate the effect of this uncertainties on the GSMF, we have performed 30 Monte Carlo realizations. First, we have scrambled the flux in each band assuming a normal distribution given by the errors in photometry. Then, photometric redshifts and stellar masses have been extracted for each source from \textsc{LePHARE}. The methodology is the same followed for estimating the Eddington Bias, as explained in Sect. \ref{subsec:method_eddington}.

Having done this, we then calculated a set of 1/Vmax points following the methodology described in Sect. \ref{subsec:method_vmax}. In panel 2 of Fig. \ref{fig:smf_uncert} we show the scatter of the 1/Vmax points, with the 30 Monte Carlo 1/Vmax points shown in gray and the real value superimposed in green. The effect is greater toward stellar masses $\mathcal{M} \lsim 10^{8.5} {\rm \ M_\odot}$, with a scatter of $0.4-0.5$ dex. It becomes milder for galaxies with stellar masses $\mathcal{M} \gtrsim 10^{8.5} {\rm \ M_\odot}$, with typical scatters of $0.1-0.2$ dex.
\subsection{SED fitting templates}
Systematic uncertainties can additionally be introduced by the templates used to fit the observed photometry and star formation histories. In \cite{marchesini_evolution_2009} a thorough analysis of the effect of uncertainties on the GSMF is made. It is found that testing a set of different set of stellar templates, IMFs and metallicities can have a non negligible effect on the realization of the GSMF.

In the present work we will only take into account random uncertainties. The high quality of the photometry we use together with the wide spectral coverage of \textit{HST} and \textit{JWST} minimize the offsets in photometric redshift and stellar mass even when using different templates for SED fitting.
\subsection{Nebular emission}
As described in Sect. \ref{sec:dataset_sedfitting}, we assesed the effect of nebular emission on the GSMF by including \texttt{SB99} stellar templates \citep{leitherer_starburst99_1999} which account for young (age$<10$ Myr) and star forming (with constant star forming history) galaxies. The final masses and photometric redshifts have been obtained from the \textsc{LePHARE} run between \texttt{BC03} and \texttt{SB99} that has a smaller reduced $\chi^2$.

In our case, about $15$\% and $18$\% of the total number of galaxies have lower reduced $\chi^2$ when using \texttt{SB99} models for HUDF and PRIMER-UDS, respectively. To test the significance of this result to $1\sigma$, we impose that the difference of reduced $\chi^2$ is greater than 1 ($\Delta \chi^2_{red} > 1$). The amount of galaxies that are better fit by \texttt{SB99} templates with $1\sigma$ significance is of $4$\% and $1$\% respectively. The effect on including nebular emission is shown in panel 3 of Fig. \ref{fig:smf_uncert}. The scatter produced by using different templates is within the error bars for all masses and redshifts, so we do not expect the inclusion of nebular lines (following this methodology) to significantly alter the shape of the GSMF.

As studied in \citet{rinaldi_strong_2023}, strong line emitters can show a flux excess in their photometry, particularly in medium bands. For assessing the effect of line contamination in the estimation of stellar masses, we re-run the SED fitting removing the possible medium bands affected by typically strong lines or lines complexes (i.e., $\rm H\alpha$ and $\rm H\beta+OIII$) on a galaxy-by-galaxy basis. After doing so, we find that the fraction of galaxies that show a difference on stellar mass greater than $0.3$dex is around $12\%$, with a mean offset of around $0.1$dex. To all effects, within the uncertainties obtained for our study, we don't expect emission lines to produce a significative effect on estimation of the stellar mass.
\subsection{Cosmic variance}\label{subsect:cosmic_variance}
The small size of the fields we use can result in a a sampling of a galaxy over or under density, creating therefore a field to field variation of the galaxy population. This effect is more important for the higher stellar mass galaxies ($\mathcal{M} \gtrsim 10^{10} {\rm \ M_\odot}$) as these objects are less common, but still can play an important role in the overall shape of the GSMF.

We have estimated the effect of the cosmic variance using the method implemented following \cite{moster_cosmic_2011}. While other methods can be found in literature (e.g., \citealp{trenti_cosmic_2008}) but the one of \cite{moster_cosmic_2011} has the advantage of considering the differential effect for a variety of field sizes, stellar masses and redshifts. The error contribution from cosmic variance can be seen in Fig. \ref{fig:err_budget}.

Since in this work we are considering to fields, HUDF and PRIMER-UDS, a direct comparison between the 1/Vmax points obtained in one and the other can be done. The right panel of Fig. \ref{fig:smf_uncert} shows the GSMF calculated using each of the fields, HUDF in blue and PRIMER-UDS in red. An overall good agreement between both is found, with a scatter contained within the error bars for the 1/Vmax points as seen in the right panel of Fig. \ref{fig:smf_uncert}.

As shown in Fig. \ref{fig:err_budget}, cosmic variance dominates the error budget for all masses and redshifts. Poissonian and photometric uncertainties (that have been computed following the method described in Sect. \ref{subsect:poisson_uncert} and Sect. \ref{subsect:photom_uncert}, respectively) are comparable in their overall contribution to the error budget. All contributions from random errors have been added in quadrature for our analysis.

\subsection{\rm Treatment of the Eddington bias} \label{subsec:method_eddington}
Uncertainties in stellar mass and photometric redshift propagate into the GSMF. The exponential decline of the Schechter function at  $\mathcal{M}>M^\ast$ further enhances this effect, which becomes asymmetric and introduces systematics into the estimation of all three Schechter function parameters $(\alpha,M^\ast,\phi^\ast)$. This is known as the Eddington Bias \citep{eddington_formula_1913}. The effect of the Eddington bias on the Schecter parameters has been explored multiple times in the literature (e.g., \citealt{caputi_stellar_2011, ilbert_mass_2013, caputi_spitzer_2015, grazian_galaxy_2015}), and more recently in \citealp{davidzon_cosmos2015_2017, obreschkow_eddingtons_2018, kokorev_evolving_2021, weaver_cosmos2020_2022}).

In \cite{ilbert_mass_2013} a product of normal and Cauchy-Lorentz kernels was convolved with the Schechter function in order to take into account the effect of the Eddington bias. The free parameters $(\sigma, \tau)$ of the kernel $\mathcal{K}(\sigma, \tau)$ are obtained from the stellar-mass error distribution. The convolution (as implemented in \citealp{ilbert_mass_2013}) is defined in Eq. \ref{eq:conv_eding} where the kernel $\mathcal{K}(\rm{M},z)$ is the one detailed in Eq. \ref{eq:ilbert_kernel}, where $\sigma$ and $\tau$ are the dispersion parameters of the Gaussian and Lorentzian error distributions, respectively.

\begin{equation}\label{eq:conv_eding}
\Phi^{conv}({\rm M}) = \int_{-\infty}^{\infty} \Phi({\rm M}|\alpha,M^\ast,\phi^\ast) \mathcal{K}({\rm M}-m,z)dm \,.
\end{equation}

\begin{equation}\label{eq:ilbert_kernel}
    \mathcal{K}({\rm M},z) =  G[{\rm M},\sigma] \times L[{\rm M},\tau] = \frac{1}{\sqrt{2\pi}\sigma} exp\left(-\frac{(\log_{10}{\rm M}-x)^2}{\sigma^2}\right) \times \frac{1}{2\pi}\frac{\tau}{[(\log_{10}{\rm M})^2-x^2] + (\frac{\tau}{2})^2} \,.
\end{equation}

In general, $\sigma$ and $\tau$ can depend on redshift and stellar mass. However, \cite{ilbert_mass_2013} found that $\sigma$ and $\tau$ could be assumed constant for all stellar-mass bins but had a mild evolution with redshift. In particular, the error distributions becoming broader with increasing redshift. However, their sample consists of lower redshift and higher stellar mass galaxies compared to ours. This is why in our case, the evolution of error distributions with stellar mass cannot be ruled out.

\cite{grazian_galaxy_2015} made use of the stellar mass probability distributions to correct for the Eddington bias. Their error distribution is not constant for all stellar masses, becoming narrower for higher mass galaxies. This is the driving factor for a milder effect at high masses and a stronger one at low masses compared to \cite{ilbert_mass_2013}, as is discussed in \cite{grazian_galaxy_2015} but also in \cite{davidzon_cosmos2015_2017}.

In this work, we make use of a modified version of the method presented in \cite{ilbert_mass_2013}. We convolve the Schechter function with stellar mass dependent kernel that is the product of a Gaussian and a standard Student-T distribution, that can be defined as:
\begin{equation}
    \mathcal{K}({\rm M}) =  G[{\rm M},\sigma,\Delta] \times T[{\rm M},df] = \frac{1}{\sqrt{2\pi}\sigma} exp\left(-\frac{(\log_{10}({\rm M}-x-\Delta)^2}{\sigma^2}\right) \times \frac{\Gamma(\frac{df+1}{2})}{\Gamma(df/2) \sqrt{\pi \, df}} \left(1+{\rm M}^2/df\right)^ {-\frac{df+1}{2}}  \,,
\end{equation}
where $\sigma$ and $df$ are the dispersion parameters of the Gaussian and Student-T error distributions, respectively, and $\Delta$ is the mean of the Gaussian component, allowing for asymmetry in the error distributions.

\begin{figure}[t]
\plottwo{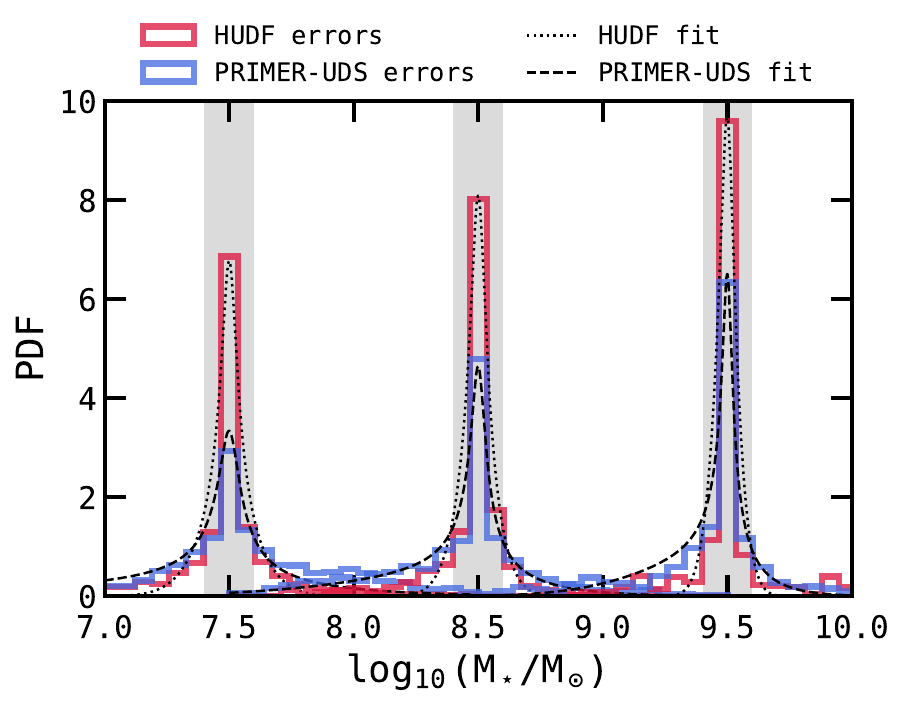}{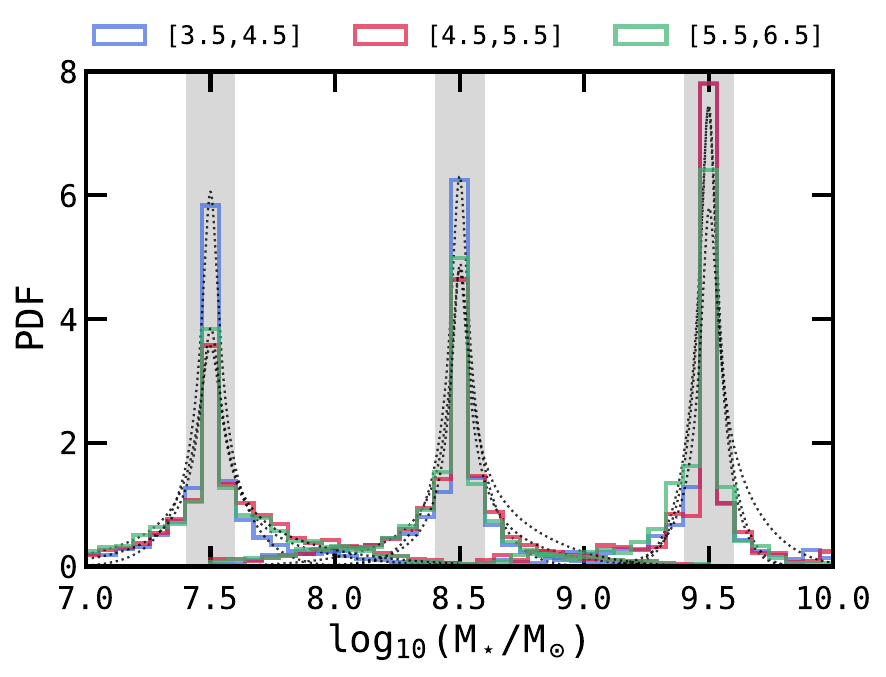}
\caption{\textbf{Left:} Stellar mass error distributions for $3.5 \leq z \leq 4.5$ galaxies, obtained after 30 realizations of stellar mass and photometric redshifts using randomized photometry. The actual error distributions for HUDF (red) and PRIMER-UDS (blue) are shown. The fits of our kernel are shown as dotted and dashed black lines, respectively. Vertical grey bands show the central value of each stellar mass bin, namely $log_{10}(\mathcal{M}/{\rm \ M_\odot}):7.5, 8.5, 9.5$. \textbf{Right:} Stellar mass error distributions for the combined PRIMER-UDS + HUDF fields. Each color shows a different redshift bin, $z:4,5,6$. The grey bands show the central value of each mass interval ($log_{10}(\mathcal{M}/{\rm \ M_\odot}) = (7.5, 8.5, 9.5)$). A clear evolution of the errors with stellar mass and redshift is present in our data.}
\label{fig:eddington_uds_xdf}
\label{fig:eddington_all}
\end{figure}

The main reasons that have motivated our reparametrization of the kernel are that we allow for non-symmetrical distributions (which can be relevant for the lowest stellar mass bins), and that the Student-T distribution provides a better fit to the actual errors of our data. In addition, we carefully study the dependance of the errors with both stellar mass and redshift.

The method that we developed can be summarized in the following steps:
\begin{enumerate}
\vspace{-2 mm}
    \item We divided the sample into stellar mass and redshift bins.
    \vspace{-2 mm}
    \item We obtained 30 realizations of stellar mass and photometric redshift after randomizing the photometry within the error bars using \textsc{LePHARE} with the same configuration as the original fitting.
    \vspace{-2 mm}
    \item We derived the stellar mass error distribution for each photometric redshift and stellar mass bin by studying the differences between the original masses and the 30 \textsc{LePHARE} realizations.
    \vspace{-2 mm}
    \item We fitted the Gaussian $\times$ Student-T kernel $\mathcal{K}(\sigma, df, \Delta)$ to the error distribution. Here $df$ are the degrees of freedom of the Student-T distribution and $\Delta$ is the skewness of Gaussian component of the error distribution.
    \vspace{-2 mm}
\end{enumerate}

Panel 1 of Fig. \ref{fig:smf_uncert} shows the Eddington bias correction effects on the shape of the GSMF for the redshift range $z:3.5 \leq z \leq 4.5$. It can be seen that our approach produces a significant effect in both low and high-mass end of the GSMF, thus altering the values of $\alpha$ and $M^\ast$.

Fig. \ref{fig:eddington_uds_xdf} shows the error distributions for three stellar-mass ranges obtained by considering 30 realizations of photometric redshift and stellar mass from \textsc{LePHARE} using as input scrambled photometry within the error bars. In the plot we only show the distributions for $3.5 \leq z \leq 4.5$, but for both fields HUDF (red) and PRIMER-UDS (blue). The Gaussian times Student-T kernel fittings are shown in dotted and dashed black lines, respectively.

Our error distributions are skewed towards lower masses. Such an effect has been also found in \cite{grazian_galaxy_2015}. We also show that the deeper data in HUDF benefits from smaller photometric uncertainties, which results in an overall smaller scatter in stellar mass and less skewness in the error distribution. Fig. \ref{fig:eddington_all} shows the error distributions for the combined HUDF + PRIMER-UDS sample. We conclude that a clear evolution of the error distribution with both stellar mass and redshift is present for the total sample. However, combining both fields results in a less skewed error distribution, especially compared to the one from PRIMER-UDS (this is a shallower dataset with larger photometric uncertainties at any given stellar mass).

\newpage
\bibliography{references}{}
\bibliographystyle{aasjournal}
\end{document}